\documentclass[useAMS,usenatbib]{mn2e}

\usepackage{graphicx}
\usepackage{amsmath}
\usepackage[dvips,usenames]{color}

\def\aj{AJ}%
\def\apj{ApJ}%
\def\apjl{ApJ}%
\def\aap{A\&A}%
\def\mnras{MNRAS}%
\def\pasp{PASP}%
\newcommand{\realfigure}[3]{
              \begin{figure}
              \includegraphics[width=84mm]{#1}
              \caption{#2}
              \label{#3}
              \end{figure}}

\newcommand{\realfiguretwofig}[4]{
              \begin{figure*}
              \centering
	          \includegraphics[width=84mm]{#1}\hfill
	          \includegraphics[width=84mm]{#2}
              \caption{#3}\label{#4}
              \end{figure*}}

\newcommand{\mytab}[6]{
\begin{table}
\caption{#4}
\label{#5}
\begin{center}
\begin{#6}
\begin{tabular}{#1}
\hline 
#2
\hline
#3
\hline
\end{tabular}
\end{#6}
\end{center}
\end{table}
}

\newcommand{\mytabbig}[6]{
\begin{table*}
\begin{minipage}{126mm} 
\caption{#4}
\label{#5}
\begin{center}
\begin{#6}
\begin{tabular}{#1}
\hline 
#2
\hline
#3
\hline
\end{tabular}
\end{#6}
\end{center}
\end{minipage}
\end{table*}
}

\newcommand{\trilegal}{{\sc trilegal}}

\newcommand{\leoi}{\mbox{Leo\,I}}
\newcommand{\leoii}{\mbox{Leo\,II}}

\newcommand{\feh}{{\rm [Fe/H]}}

\newcommand{\mh}{{\rm [M/H]}}

\newcommand{\corrmod}{$(m-M)_0=22.04\pm 0.11$}
\newcommand{\corrmh}{[M/H]\,$=-1.24\pm 0.05\,(\text{int})\pm 0.15\,(\text{syst})$}

\newcommand{\gaumhval}{-1.45}
\newcommand{\gaumhdisp}{0.19}
\newcommand{\mhval}{-1.51}
\newcommand{\mhcorr}{-1.24}

\newcommand{\numc}{37}
\newcommand{\numo}{39}

\newcommand{\kk}{K}
\newcommand{\Mks}{$M_{K}$}
\newcommand{\vks}{\mbox{$V\!-\!K$}}
\newcommand{\hks}{\mbox{$H\!-\!K$}}
\newcommand{\jks}{\mbox{$J\!-\!K$}}

\newcommand{\jh}{\mbox{$J\!-\!H$}}

\newcommand{\ebv}{\mbox{$E_{B\!-\!V}$}}

\newcommand{\Msun}{\mbox{$M_{\odot}$}}

\newcommand{\caii}{Ca\,{\sc ii}}
\def\nodata{~--}

\newcommand{\referee}[1]{#1}         

\newcommand{\tbd}[1]{}  

\newcommand{\abbrev}[1]{{#1}}      

\title[A near-infrared study of AGB and red giant stars in the \leoi\
dSph]{ A near-infrared study of AGB and red giant stars in the \leoi\
  dSph galaxy}

\author[Held et al.]{
E. V. Held$^{1}$\thanks{E-mail: enrico.held@oapd.inaf.it},
M. Gullieuszik$^{1}$,
L.~Rizzi$^{2}$,
L.~Girardi$^{1}$,
P.~Marigo$^{3}$ and
I.~Saviane$^{4}$
\\
$^{1}$INAF/Osservatorio Astronomico di Padova,
vicolo dell'Osservatorio 5, I-35122 Padova, Italy
\\
$^{2}$Joint Astronomy Centre, 660 N. A'ohoku Place, 
University Park, Hilo, HI 96720, USA
\\
$^{3}$Dipartimento di Astronomia, Universit\`a di Padova, 
vicolo dell'Osservatorio 2, I-35122 Padova, Italy
\\
$^{4}$ European Southern Observatory, 
Casilla 19001, Santiago 19, Chile
}

\begin{document}

\date{Accepted \dots Received \dots; in original form \dots}

\pagerange{\pageref{firstpage}--\pageref{lastpage}} \pubyear{2008}

\maketitle

\label{firstpage}

\begin{abstract}
  A near-infrared imaging study of the evolved stellar populations in
  the dwarf spheroidal galaxy \leoi\ is presented.  Based on $JH \kk$
  observations obtained with the WFCAM wide-field array at the UKIRT
  telescope, we build a near-infrared photometric catalogue of red
  giant branch (RGB) and asymptotic giant branch (AGB) stars
  in \leoi\ over a $13\farcm5$ square area.
  The \vks\ colours of RGB stars, obtained by combining the new data
  with existing optical observations, allow us to derive a
  distribution of global metallicity \mh\ with average $\mh =
  \mhval$ (uncorrected) or \corrmh\ after correction for the mean age
  of \leoi\ stars. This is consistent with the results from
  spectroscopy once stellar ages are taken into account.
  Using a near-infrared two-colour diagram, we discriminate 
  between carbon- and oxygen-rich AGB stars and obtain a clean
  separation from Milky Way foreground stars. 
  We reveal a concentration of C-type AGB stars relative to the red
  giant stars in the inner region of the galaxy, which implies a
  radial gradient in the intermediate-age (1--3 Gyr) stellar
  populations.
  The numbers and luminosities of the observed carbon- and oxygen-rich
  AGB stars are compared with those predicted by evolutionary models
  including the thermally-pulsing AGB phase, to provide
  new constraints to the models for low-metallicity stars.
  We find an excess in the predicted number of C stars fainter than
  the RGB tip, associated to a paucity of brighter ones. The number of
  O-rich AGB stars is roughly consistent with the models, yet their
  predicted luminosity function is extended to brighter luminosity.
  Although these discrepancies can be partly ascribed to significant
  uncertainties in the \leoi\ star-formation history and
  incompleteness of the spectroscopic samples of C stars fainter than
  the RGB tip, it appears more likely that the adopted evolutionary
  models overestimate the C star lifetime and underestimate their
  $\kk$-band luminosity.  

\end{abstract}

\begin{keywords}
Galaxies: individual: \leoi\ -- 
Galaxies: stellar content --
stars: AGB and post-AGB -- 
stars: carbon -- 
Local Group
\end{keywords}

\section{Introduction}

The dwarf spheroidal galaxy \leoi\ was the first shown to be dominated
by a relatively young stellar population \citep{lee+1993}.  HST deep
photometry confirmed the presence of a prominent intermediate-age
stellar population and young stars \citep{gall+1999,capu+1999}.
Quantitative analysis of the HST data established that most star
formation activity (70\%--80\%) occurred between 7 and 1 Gyr ago, with
some residual star formation until at least $\sim 300$ Myr ago.
The detection of a blue horizontal branch \citep{held+2000} provided
the first evidence for an old stellar population, later confirmed by
the discovery of RR Lyrae variable stars \citep{held+2001}.

The presence of an intermediate age population in \leoi\ was also
revealed by an extended asymptotic giant branch (\abbrev{AGB}).  After
the spectroscopic discovery of a carbon star in \leoi\ by
\citet{aaro+1983}, $JHK$ follow-up photometry by \citet{aaromoul1985}
showed that the carbon stars (4 newly identified on the basis of their
red near-infrared colours) belong to an extended \abbrev{AGB} typical
of an intermediate-age stellar population.
The presence of a significant population of C stars (19 stars) in
\leoi\ was confirmed by objective grating spectroscopy in the
blue-green spectral region (Swan C$_2$ bands) by
\citet{azzo+1985,azzo+1986}.
Using a 4-filter technique based on the $RI$ filters and CN and TiO
intermediate-band filters, \citet{demebatt2002} further surveyed
\leoi\ for C stars, identifying 13 C-star candidates, 6 of which in
common with \citet{azzo+1986}.
\citet{menz+2002} obtained near-infrared (\abbrev{NIR}) 
photometry to $K_\text{s}=16$
in a $7\farcm2$ square field allowing a study of the AGB content of
\leoi.  They showed that most luminous AGB stars are carbon stars,
several of which are variable, and detected the presence of 5 very red
stars near the AGB tip, probably young ($\sim 2$ Gyr old)
dust-enshrouded stars, also candidates for Mira variability.

Several spectroscopic studies recently addressed the metallicity
and chemical evolution of \leoi.  \citet{bosl+2007} 
used the \caii\ triplet method to analyse
Keck-LRIS spectra of 102 red giant branch 
(\abbrev{RGB}) stars, and found ${\rm [Fe/H]}=-1.34$.  A
mean metallicity \feh\ $=-1.31$ was measured by \citet{koch+2007leo1}
for 58 red giants.  In both studies, the metallicity distribution is
well described by a Gaussian function with a 1$\sigma$ width of 0.25
dex, and a full range in \feh\ of approximately 1 dex.
In \citet{gull+2009spec} we obtained metallicities for 54 stars in \leoi\
from measurements of the Ca triplet line strengths, and found a mean
metallicity $\feh=-1.4$ and \mh\ $=-1.2$ on a scale tied to the \mh\
ranking of Galactic globular clusters (\abbrev{GGCs}). 
A narrow and symmetric metallicity distribution, with a very low
instrinsic metallicity dispersion, $\sigma_{\rm [M/H]} = 0.08$ dex,
appears 
consistent with a prompt initial enrichment and galactic winds
expelling the metals products of stellar evolution.
We also suggested the presence of a radial age gradient, with RGB
stars in the inner part of \leoi\ being on average younger than those
in the outer regions.

In this study, we further investigate the star-formation history of
\leoi\ and the presence of stellar population gradients using deep
\abbrev{NIR} wide-field observations and analysing the
AGB stars as tracers of intermediate-age stellar populations.
Our NIR observations provide new constraints on the star-formation
history of the galaxy, which are set by the number, type (carbon
vs. oxygen-rich), and luminosities of AGB stars \citep[see, 
e.g.,][]{cion2007,groe2007}.
Near-infrared photometry is ideal to this purpose because the cool AGB
stars emit most of their flux at near-infrared wavelengths.  The most
evolved, dust-enshrouded AGB stars can be detected only at infrared
wavelengths.  In addition, the NIR photometry of red giant stars can
be used to provide independent information on basic stellar
properties.

Together with a companion paper on \leoii\ \citep{gull+2008leo2IR},
this study is part of a survey of evolved stellar populations in dwarf
galaxies in the Local Group and its vicinity, aimed at providing an
observational database for the calibration of AGB stars models in a
wide range of metallicities as a tool for a reliable interpretation of
observations.  For distant resolved systems (e.g., Virgo cluster
galaxies observed with JWST or ELT) this will represent the main
observational fact for modelling the galaxy's star formation rate in the
last few Gyr.

\section{Observations and data reduction}
\label{s:leo1obs}

\mytab{
cccccc}{
Filter&
$N_{\rmn{ima}}$&
DIT(s)&
$N_{\rmn{exp}}$&
$N_{\rmn{jit}}$&
Microsteps\\}{
$J$  & 6 & 5.0 & 2 & 9 & $2\times2$ \\
$H$  & 6 & 5.0 & 2 & 9 & $2\times2$\\
$\kk$  & 10 & 5.0 & 2 & 9 & $2\times2$\\}{
Observing log.
}{
t:obslog}{
normalsize}

Observations of \leoi\ were carried out on April 19--20, 2005 using
the NIR wide field camera WFCAM \citep{casa+2007} at the UKIRT at
Mauna Kea, Hawai'i.  WFCAM is equipped with 4 Rockwell Hawaii-II
HgCdTe detectors with a scale of $0\farcs4$ pixel$^{-1}$. The size of
each detector is $2048\times2048$ pixels, corresponding to
$13\farcm6\times13\farcm6$, with a space between the arrays of 94\% of
the detector size.
  Since the tidal radius of \leoi\ is $12\farcm6$
  \citep{irwihatz1995}, one array (No.~2) was sufficient to study
  \leoi. The remaining detectors provide a useful estimate of the
  background/foreground field contamination.

  The observing strategy and instrumental setup are the same as used
  for our study of \leoii\ \citep{gull+2008leo2IR}.  The observations,
  summarised in Table~\ref{t:obslog}, consisted of 4 micro-stepped
  10\,s images (two 5\,s coadds) on a 9-points jitter pattern,
  yielding a 360\,s exposure time.  In total, 6 observations were
  obtained in $J$ and $H$ and 10 in the $\kk$ band, yielding on-target
  integration times of 36\,min in $J$ and $H$ and 60\,min in $\kk$.
  The $2\times2$ ``small'' microstepping was used to obtain a better
  spatial sampling.
\begin{table}
  \caption{
    The NIR catalogue of \leoi\ stars over WFCAM array No.~2. A
    few lines are shown here for guidance regarding its form and content,
    while the full catalogue is available from the electronic edition of the
    journal.}
\label{t:leo1catal}
\begin{center}
\begin{tabular}
{l @{\hspace{3.5mm}}c @{\hspace{3.5mm}}c @{\hspace{3.5mm}}c @{\hspace{3.5mm}}c @{\hspace{3.5mm}}c }
\hline
ID & $\alpha$ (J2000) & $\delta$ (J2000) & $J$ & $H$ & $\kk$ \\
\hline                    
14476	& 10:08:29.36	& +12:12:05.6	  & 19.48     & 18.60     & 18.43   \\  
14447	& 10:08:03.90	& +12:12:06.8	  & 19.64     & 18.93     & 18.93   \\ 
200324	& 10:08:54.77	& +12:12:06.9	  & 18.78     & 18.26     & 18.32   \\
14467	& 10:08:41.75	& +12:12:07.3	  & 20.78     & 19.95     & 20.05   \\
22945	& 10:08:38.14	& +12:12:07.7	  & 20.64     & 19.85     & 20.73   \\
\hline                  
\end{tabular}
\end{center}
\end{table}
\begin{figure*}
\begin{center}
\begin{tabular}{c c}
 \includegraphics[clip,width=84 mm]{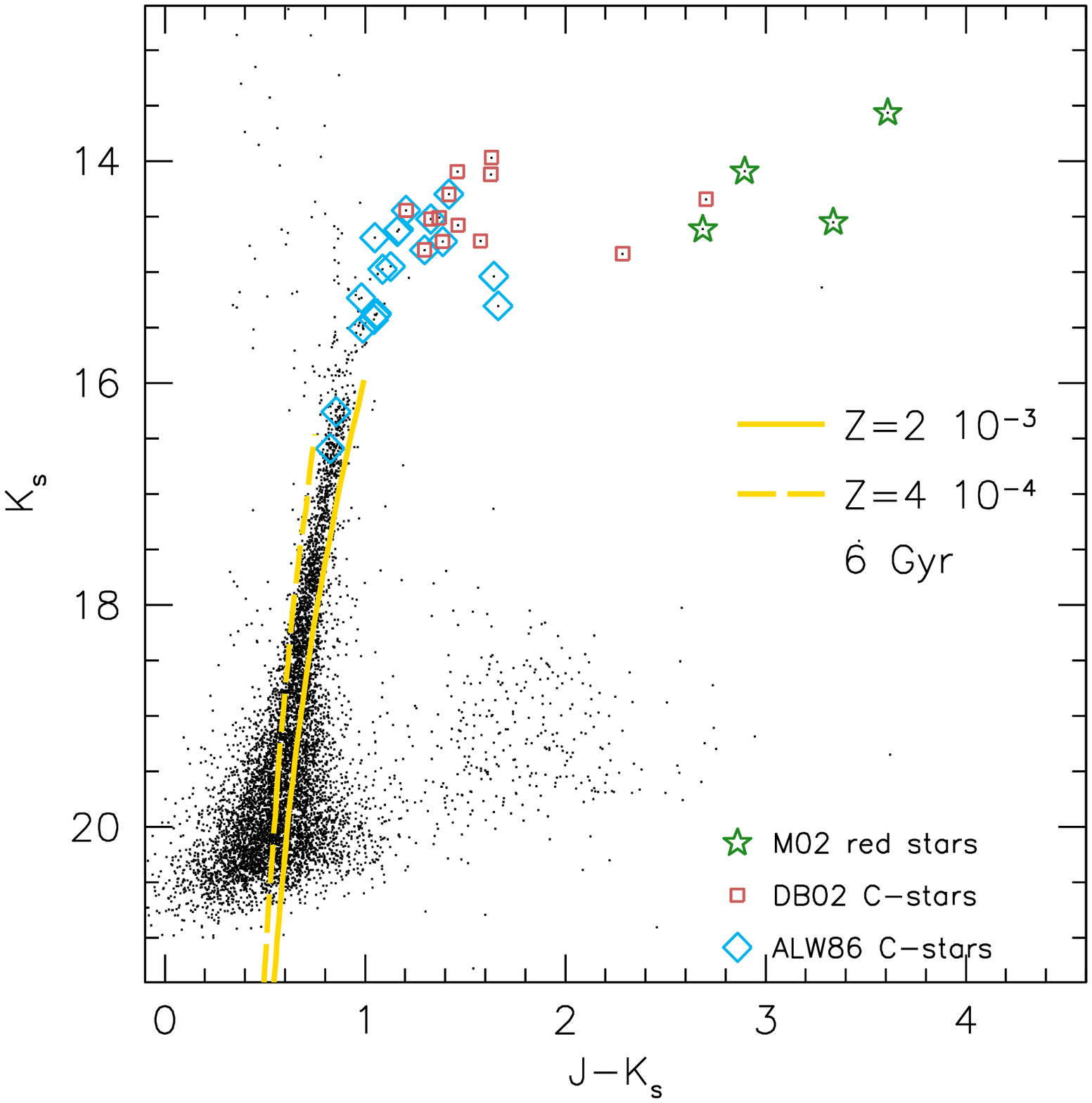}&
 \includegraphics[clip,width=84 mm]{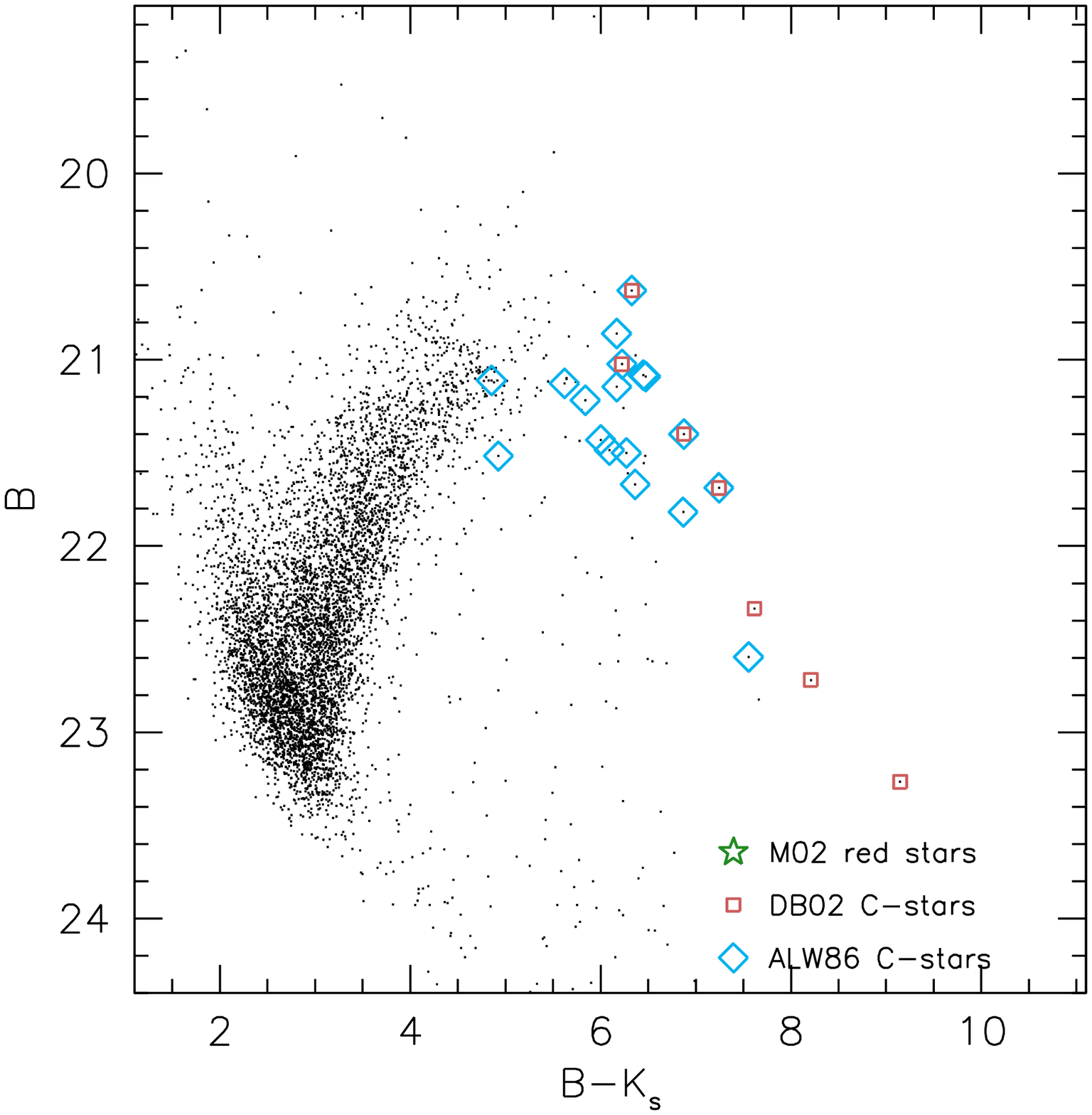}\\
\end{tabular}
\end{center}
\caption{NIR and optical-NIR colour-magnitude diagrams of
  \leoi. Superimposed on the (\jks), $\kk$\ CMD are theoretical
  isochrones from \citet{gira+2002}.  The age is 6 Gyr and the
  metallicities \referee{bracket that of \leoi: $Z=0.002$ and
  $Z=0.0004$.}  Carbon stars identified by \citet{azzo+1986} and
  \citet{demebatt2002} are shown as {\it diamonds} and {\it squares},
  respectively.  The {\it starred symbols} are the very red stars
  identified by \citet{menz+2002}.
  \label{f:4cmd} }
\end{figure*}
The raw data were calibrated using the WFCAM pipeline provided
  by the VISTA Data Flow System Project \citep{dye+2006}.  The product
  of the pipeline are 4k$\times$4k oversampled combined images
  (``leavstacks'') with a spatial resolution $0\farcs2$ pixel$^{-1}$.

  Our analysis was limited to the array centred on \leoi\ (No.~2) and
  a second array (No.~3) used as a control field to estimate the
  contamination by foreground Galactic stars and background galaxies.
  Object detection and point-spread function (\abbrev{PSF}) photometry
  was performed on each ``leavstack'' image using {\sc daophot} and
  {\sc allframe} \citep{stet1987,stet1994}. The PSF was generated with
  a ``Penny" function with quadratic dependence on the position on the
  frame.
  The astrometric calibration was performed by the pipeline using the
  2MASS Point Source Catalogue \citep[PSC]{skru+2006} as a reference,
  with a final absolute systematic accuracy of the order 
  $0\farcs1$.  The positions of the sources in the final $JH \kk$
  catalogue were converted from pixel units to the J2000 equatorial
  system using this astrometric calibration 
  and an updated version of the IRAF\footnote{The Image Reduction and
    Analysis Facility (IRAF) software is provided by the National
    Optical Astronomy Observatories (NOAO), which is operated by the
    Association of Universities for Research in Astronomy (AURA),
    Inc., under contract to the National Science Foundation.} package 
  that included support for the ZPN projection adopted by the pipeline.

  Our raw photometric catalogue was transformed to the 2MASS
  photometric system by applying the colour terms between WFCAM and
  2MASS systems derived for the UKIDSS surveys \citep{dye+2006}, and
  zero points directly computed from the median of the $J H \kk$
  magnitude differences between 2MASS and our WFCAM photometry 
  \referee{in the colour range $0.0 <$\jks$< 1.0$}, for
  stars in common with the 2MASS/PSC.  Only stars having 2MASS S/N
  ratio $> 10$ and photometric errors $< 0.1$ mag were used, which
  essentially limits the comparison to stars brighter than the
  \abbrev{RGB} tip. The r.m.s. of the residuals is 0.04 mag in the
  $J$ band and 0.05 mag in the $H$ and $\kk$ bands.
  We verified that the zero points would not change significantly ($>
  0.01$ mag) using the colour terms from \citet{hodg+2009}.  We note
  that the colour terms were defined 
  in the colour range $0.0 <$\jks$< 1.0$.  
  Outside this interval (e.g., for very red stars), our
  measurements should be considered on a (linearly transformed) WFCAM
  system rather than on the 2MASS system.
The final, calibrated NIR photometric catalogue, which includes
star-like objects detected in at least 2 images in 2 bands, is
published in the electronic issue of the journal.  A few lines are
presented in Table~\ref{t:leo1catal} to illustrate its content.

For the completeness and photometric errors we assumed the results of
artificial stars experiments used for \leoii\ \citep{gull+2008leo2IR},
since the observing strategy, exposure times, and sky conditions were
identical for the two galaxies, and the crowding is very similar and
low in the NIR.
The completeness factor is $> 50$\% for stars brighter than $\kk \sim
20$. All results of our analysis are based on photometry of stars
brighter than $\kk =18$, for which we have completeness factor $\simeq
100\%$ and internal photometric errors $< 0.02$ mag.

\section{Colour-Magnitude diagrams}\label{s:leo1cmd}

\realfiguretwofig{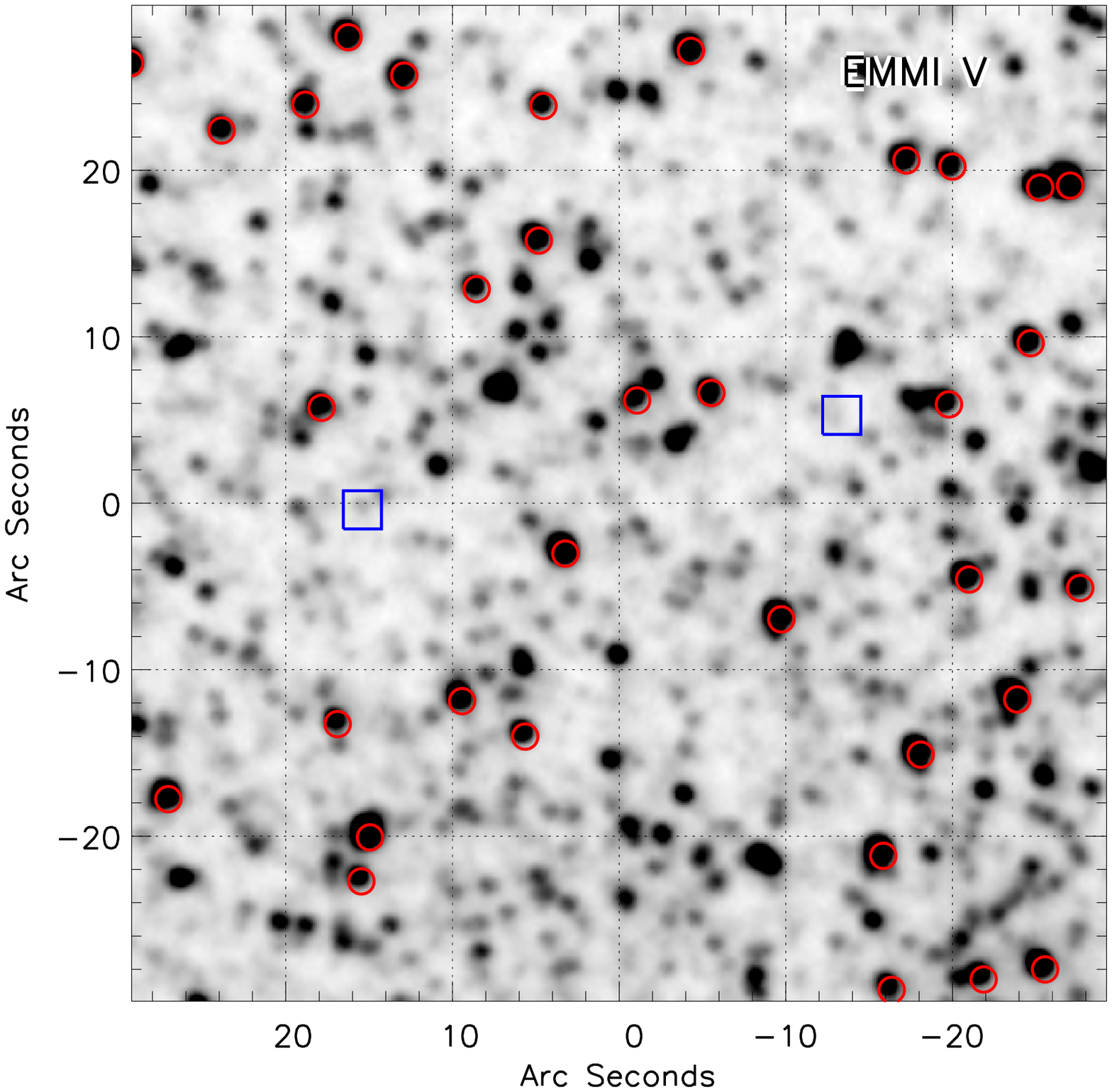}{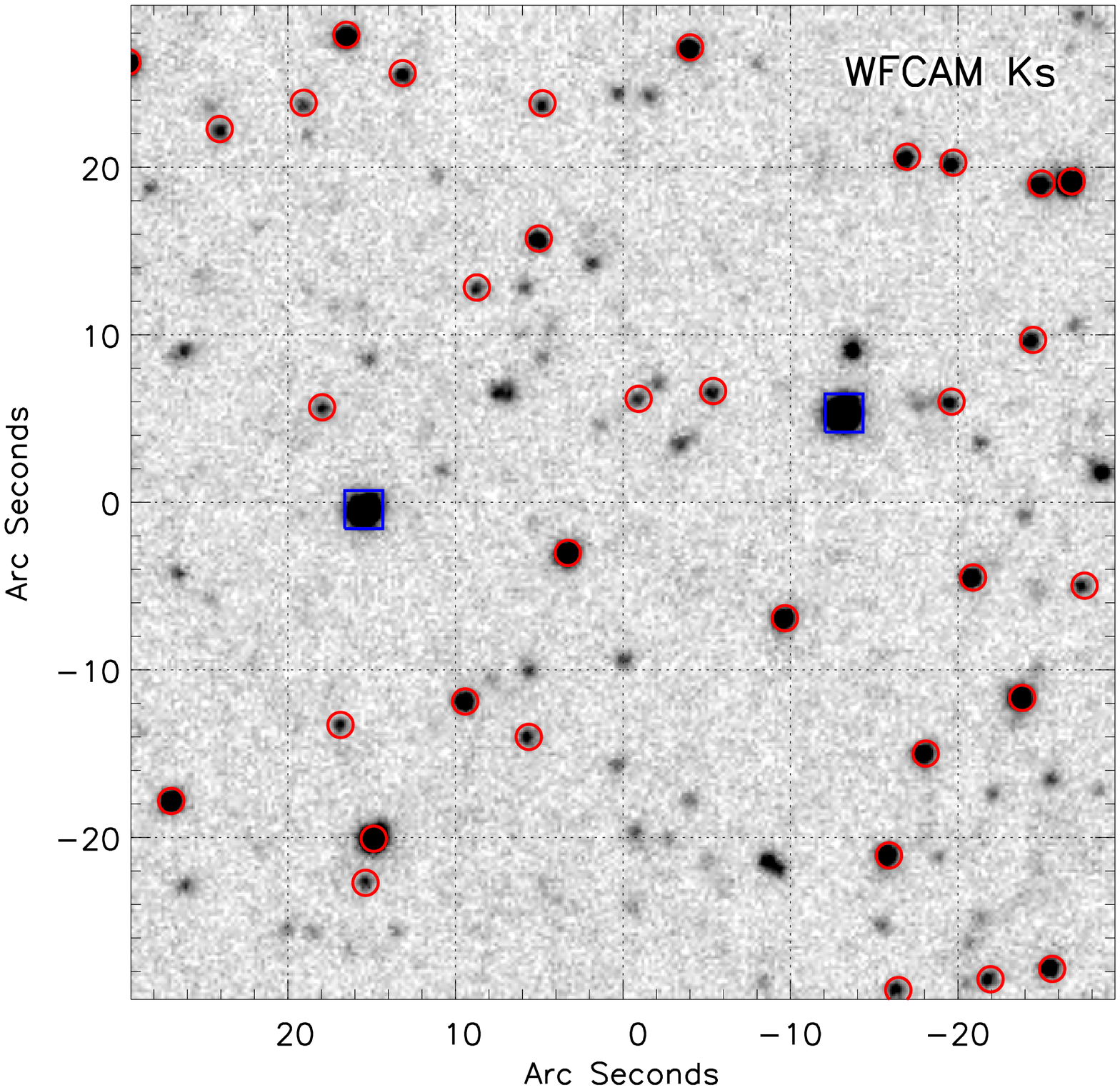}{
  A $1\arcmin\times1\arcmin$ region in \leoi\ as seen in the $V$ band
  by (NTT/EMMI data, {\it left panel}) and in the $\kk$\ band (WFCAM
  data, {\it right panel}).  {\it Open circles} mark stars brighter
  than $\kk =18$, while {\it open squares} show the position of the
  stars ``A'' (on the left side of the image) and ``B'' (on the right)
  identified by \citet{menz+2002}.  These stars are very bright in the
  near infrared and yet undetected at optical wavelengths.
}{f:mapM02}

  Figure~\ref{f:4cmd} presents our (\jks), $\kk$ colour-magnitude
  diagram (\abbrev{CMD}) of the \leoi\ field, along with a diagram
  combining NIR data and optical photometry obtained with the EMMI
  camera at the ESO/La Silla NTT telescope \citep{held+2000}.
  Our photometric catalogue was cleaned using the {\it sharp}
  parameter calculated by {\sc allframe} to remove noise peaks,
  diffuse objects, and other spurious detections. 
  \referee{ A magnitude-dependent {\it sharp} cutoff was adopted,
    as in \citet[][their figure 2]{gull+2007sag}.}

  Our dataset provides a nearly complete sample of upper-AGB stars
  belonging to the intermediate-age stellar populations of \leoi, with
  a red tail of C stars which extends to $(\jks) \sim 3.5$, in
  agreement with \citet{menz+2002}.
  In addition, the NIR CMD of \leoi\ samples the RGB from
  the tip (clearly visible at $\kk \sim 16$) down to $\kk \sim
  20.5$. The RGB is comprised between the \citet{gira+2002}
  theoretical isochrones for an age 6 Gyr and 
  \referee{metallicities $Z=0.002$ and $Z=0.0004$.}

  \referee{ The cool AGB C-stars (mostly variable) are the brightest
    objects in \leoi\ in the $\kk$ band.  At optical wavelengths,
    their luminosity dramatically decreases and the reddest among them
    are generally undetected in the $B$ band.  The completeness of the
    optical catalogues is also complicated by the large amplitude of
    luminosity variations in the optical domain.  Still, many
    upper-AGB stars are present in the optical catalogues, although
    some are very faint (see Fig.~\ref{f:4cmd}). As a result, archival
    optical catalogues may be used to retrieve AGB stars detected in
    the NIR (or mid-infrared) and characterise their SEDs.  }
  We cross-identified the stars in our NIR catalogue with the carbon
  star surveys of \citet{azzo+1986} and \citet{demebatt2002} and with
  the list of very red objects in \citet{menz+2002}, using positional
  coincidence within a $2\farcs 5$ radius.
  Star C10 was incorrectly identified by \citet{demebatt2002} with
  star no.~7 in the \citet{azzo+1986} catalogue. Since the distance
  between D02-C10 and ALW-7 is $32\arcsec$, and both stars are located
  within $2\farcs 5$ from two distinct sources in our catalogue (with
  colours and magnitudes typical of C stars), we conclude that they
  are distinct stars.
  All of the very red stars detected by \citet{menz+2002} within our
  field-of-view (namely, stars A, B, C, and D) are also present in our
  NIR CMD (Fig.~\ref{f:4cmd}). Among them, only star D is
  visible in the optical bands. As an example, Fig.~\ref{f:mapM02}
  shows two of the reddest stars. They are very bright in the $\kk$
  band, yet fall below the detection threshold in the $V$ image.
 
  Different techniques adopted for searching C stars lead to samples
  with different biases \citep[see ][]{batt+deme2009}.  The reddest
  stars are too faint in the optical spectral range to be detected
  with objective grating techniques, while the 4-filter technique
  misses the bluest stars \citep{demebatt2002}.
  \citet{azzo+1986}, using an objective grating operating in the
  $5000-7000$ \AA \ range, did not detect the brightest, reddest C
  stars \citep[as noticed by ][]{menz+2002}, while
  \citet{demebatt2002} could not retrieve the bluest, i.e.  warmest,
  spectroscopically identified C stars.

  In Sect.~\ref{s:leo1-AGB} we will present a method for selecting C-
  and O-rich AGB stars based on two-colour NIR diagrams. Our selection
  will include all previously known C-stars and, in addition, will
  allows us to separate the foreground Milky Way stellar population.

\section{Distance}\label{s:leo1dist}

  The luminosity of the RGB tip (\abbrev{TRGB}) in the $I$ band has
  long been used as a valuable distance indicator 
  \citep{dacoarma1990,lee+1993}.  The dependence of the TRGB
  luminosity on age and metallicity is minimised in the $I$-band
  \citep[e.g.,][]{salagira2005}.  In the NIR, the luminosity of the
  tip depends on age and metallicity in a more complex way. For
  instance, the $\kk$-band magnitude of intermediate-age stars at the
  TRGB is {\it fainter} than that of old stars; while the TRGB $\kk$
  luminosity rises with increasing metallicity
  \citep[e.g.,][]{salagira2005}.  For a population that becomes more
  metal-rich with time as a result of galaxy chemical evolution, the
  two effects may partly balance.
  As discussed in \citet{gull+2007for,gull+2008leo2IR},
  distance estimates consistent with those obtained from optical
  methods can be derived from NIR photometry, provided that 
  the star-formation history (\abbrev{SFH}) of
  the galaxy is at least approximately known. Testing our ability to
  measure distances from NIR data is important in view of
  the future generation of ground based and space telescopes that 
  will operate mainly or exclusively at infrared wavelengths (JWST,
  adaptive optics at the ELT).

  In a complex stellar population, the RGB is generally wide and the
  cutoff does not have a constant luminosity \citep[see, e.g., the
  discussion of Fornax dSph by][]{whit+2009}.  However, the TRGB can
  still be operationally defined to represent the mix population, and
  compared with synthetic CMDs that reflect the \abbrev{SFH} of the
  system.
  The method is employed here to measure the distance to \leoi.  An
  objective estimate of the magnitude of the TRGB was obtained by
  fitting the $\kk$-band luminosity function with a step function
  convolved with a Gaussian kernel representative of the photometric
  errors.  
  This method, extensively applied by our group
  \citep[e.g.,][]{moma+2002}, has given in the $I$ band results
  consistent within $1 \sigma$ with the Maximum Likelihood Algorithm
  of \citet{maka+2006} \citep[see ][]{rizz+2007for}.

  The magnitudes of the TRGB measured in the $JH\kk$ bands are listed
  in Table~\ref{t:trgb}.  The errors associated to these magnitudes
  are essentially the uncertainty on the absolute photometric
  calibration, since the uncertainties due to the fitting algorithm
  are less than 0.01 mag and the internal photometric errors at the
  level of the TRGB are negligible (see Sect.~\ref{s:leo1obs}).
  To derive the distance to \leoi, the $JH \kk$ TRGB magnitudes were
  compared with the empirical calibrations of $M_\lambda^{\rm TRGB}$
  as a function of \mh\ from \citet{vale+2004b}.  
  This calibration, whose intrinsic systematic error is $\sigma=0.16$
  mag, is based on Galactic globular clusters, i.e. old stellar
  populations.  Applying it to the complex stellar populations of
  \leoi\ requires a ``population correction'' based on synthetic CMDs
  to take into account its star formation history.
  The distance to \leoi\ was then computed as 
  \begin{equation}\label{e:trgbmM}
    (m-M)_0 = m_\lambda - A_\lambda - (M_\lambda + \Delta M_\lambda)
  \end{equation}
  where $m_\lambda$ is the observed TRGB magnitude, $A_\lambda$ is the
  extinction, $M_\lambda$ is the absolute TRGB magnitude as derived
  from \citet{vale+2004b} calibration, and
  $\Delta M_\lambda$ is the population correction needed to account
  for the effects of the different 
  \abbrev{SFH} of \leoi\ and the \abbrev{GGCs}.

\mytab{
lccccc}{
band&
$m^{\rm TRGB}$&
$\hspace{5pt}M^{\rm all}$&
$\hspace{5pt}M^{\rm old}$&
$\mu_0$&
$\sigma_{\mu_0}$\\}{
$J$		&17.00	&      $-5.310$ &        $-5.326$& 22.09 &0.18\\
$H$		&16.30	&      $-6.179$ &        $-6.329$& 22.01 &0.19\\
$\kk$           &16.14	&      $-6.293$ &        $-6.340$& 22.01 &0.20\\
}{
  Observed magnitude of the TRGB and corrected distance modulus
  $\mu_0$ derived for \leoi\ from $JH\kk$\ photometry.  The distance
  was population-corrected by comparing the absolute magnitudes of the
  TRGB calculated from synthetic CMDs for the population mix of \leoi\
  ($M^\text{all}$) and a pure old population ($M^\text{old}$) (see
  text for details). 
}{
t:trgb}{
normalsize}

  The population correction $\Delta M_\lambda$ was calculated using a
  simulated CMD of \leoi\ with constant metallicity to 
  build the $JH \kk$ luminosity function (\abbrev{LF}) 
  and measure the theoretical TRGB magnitude,
  $M_\lambda^{\rm all}$ (see \citealt{gull+2007for} for details).  Our
  measurements were then repeated by selecting only stars older than
  10 Gyr ($M_\lambda^{\rm old}$).  The population correction was then
  $\Delta M_\lambda = M_\lambda^{\rm all} - M_\lambda^{\rm old}$.
  The values are listed in Table~\ref{t:trgb}.  The population
  effect is positive for all bands, i.e.  the RGB tip for the overall
  stellar population is slightly {\it fainter} than that of an old
  population.

  Our distance estimates for \leoi, computed from Eq.~\eqref{e:trgbmM}
  with a reddening \ebv~$=0.03$, are given in Table~\ref{t:trgb}.  The
  weighted mean of the distances derived from the $JH\kk$ bands is
  22.04 mag (corresponding to 256 kpc) with a standard deviation of
  0.11 mag.
  The errors were calculated by error propagation from the photometric
  calibration uncertainty (see Sect.~\ref{s:leo1obs}) and the
  intrinsic systematic error on the \cite{vale+2004b} calibration
  (0.16 mag).  
  This estimate agrees with the results of optical studies.
  \citet{held+2001} found $(m-M)_0=22.04 \pm 0.14$ from the $V$
  magnitude of \leoi\ RR-Lyrae stars; \citet{bella+2004} measured
  $(m-M)_0=22.02 \pm 0.13$ from the $I$ magnitude of the TRGB. Using
  the same method, \citet{mend+2002} found $(m-M)_0=22.05 \pm 0.10 \
  {\rm (internal)} \pm 0.18 \ {\rm (systematic)}$.  Our result
  therefore confirms the possibility of reliably estimating the distance
  to resolved stellar systems with complex stellar populations using
  NIR photometry alone.

\section{Metallicity}
\label{s:leo1metallicity}

\subsection{Metallicity distribution}

\realfigure{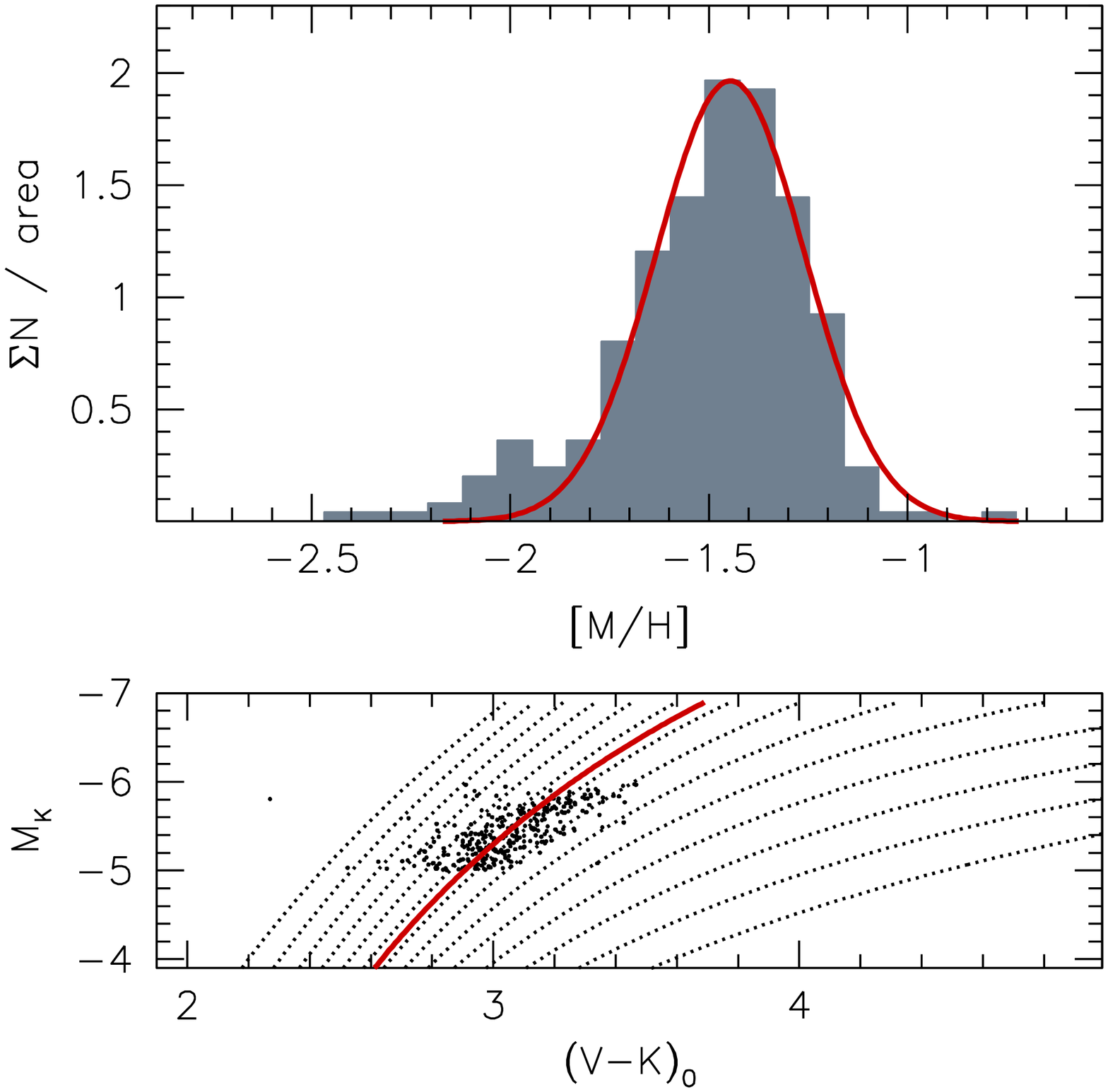}{
Derivation of a photometric Metallicity
Distribution Function of RGB stars in \leoi\ using \vks\ colours.
{\it{Upper panel:}} the metallicity distribution 
of red giants in \leoi. The {\it
solid line} is a fit of a Gaussian function to the data 
\referee{with \mh~$> -1.8$.}
The method is illustrated in the {\it{lower panel}}: the stars on the
upper RGB of Fornax are interpolated across analytical fits ({\it dotted
lines}) to the RGB fiducial lines of template Galactic globular
clusters. The {\it thick line} marks the best-fit mean metallicity.
}{f:mdf}

\realfigure{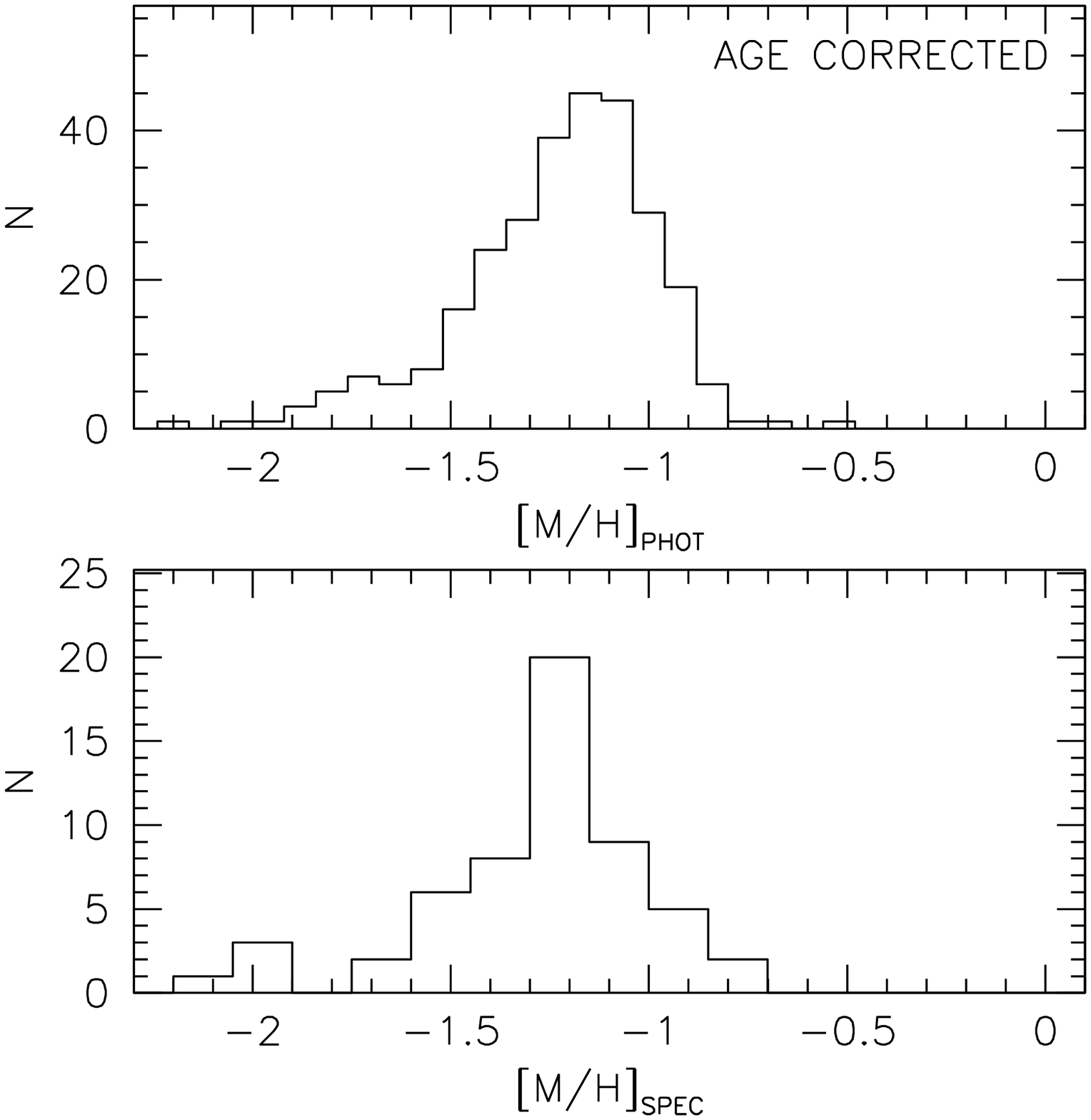}{
  Comparison of the MDFs of \leoi\ RGB stars derived from NIR
  photometry (this work) and spectroscopy \citep{gull+2009spec}.  The
  photometric MDF was corrected for age effects assuming a mean age of
  5 Gyr.
}{f:compspec1}

  The metallicity distribution function (\abbrev{MDF}) of RGB stars is
  an important constraint for models of chemical evolution of dwarf
  galaxies.  We derived a photometric MDF for RGB stars in \leoi\
  by comparing their \vks\ with the colour of GGCs of known
  metallicity, following a technique developed for optical colours
  by \citet{savi+2000b}.
  Stars on the upper RGB were interpolated in the (\vks), $M_K$
  diagram across analytical fits to the RGB fiducial lines of template
  Galactic globular clusters.  We assumed for \leoi\ a distance of
  $(m-M)_0=22.04$ and a reddening $\ebv = 0.03$.
  In our implementation of the method, we employed \mh\ values for a
  set of GGCs spanning a wide range in metallicity, from
  \citet{vale+2004a}.  The complete list of GGCs with metallicities is
  given by \citet{gull+2007for}.
  The global metallicity \mh, which measures the abundance of all
  heavy elements, is the most appropriate to estimate the
  metallicities of dwarf spheroidal galaxies (having [$\alpha$/Fe]
  ratios close to solar) by comparison with the photometric properties
  of Milky Way globular clusters, which generally show an
  overabundance of $\alpha$-elements relative to iron that is a
  function of the cluster metallicity \citep[see][ and refs.
  therein]{geis+2007}.

  The resulting {\it photometric MDF} of \leoi, obtained for red giant
  stars 
  \referee{down to $\sim 1$ mag below the TRGB ($-6.0 <$\,\Mks\,$< -5.0$), 
    is presented in
    Fig.~\ref{f:mdf} (upper panel).  With our choice, the MDF shows
    negligible contamination by stars on the blue side of the RGB,
    which probably include both younger stars in \leoi\ and Galactic
    foreground.
  The distribution has an average \mh~$= \mhval$ and is quite well
  described by a Gaussian centred at \mh~$= \gaumhval$ with a measured
  dispersion $\gaumhdisp$ dex. The standard deviation of the mean, 0.05 dex,
  is adopted as internal error of the mean metallicity.
}

  Since the stellar populations in \leoi\ are on average younger that
  the Galactic globular clusters, we computed a correction to the mean
  metallicity of \leoi\ by adopting the simplified approach that all
  RGB stars have an age of 5 Gyr \citep[e.g., ][]{gall+1999}.  Given
  two RGB stars with the same metallicity, the younger has bluer
  colours.  By interpolating over the \citet{gira+2002} isochrone set,
  we found that at $M_K=-5.0$ a star of 5 Gyr is more metal rich by
  0.27 dex than a 12.5 Gyr old star having \mh~$ = \mhval$ and the
  same \vks\ colour.
  \referee{By applying this correction, the mean metallicity of \leoi\
    from NIR photometry is then \mh~$=\mhcorr \pm 0.05$ (random) $ \pm
    0.15$ (systematic).}  The systematic error was estimated from the
  uncertainty on the photometric zero point of our calibration, which
  is 0.07 mag on the \vks\ colour (the quadratic sum of the $V$ and
  $\kk$ zero-point uncertainties).  Shifting the \leoi\ RGB by $\pm
  0.07$ mag in colour results in a $\pm 0.15$ dex variation in
  metallicity.

  The mean metallicity of red giant stars in \leoi\ in our
  spectroscopic analysis is \mh\ $=-1.26\pm 0.16$
  \citep{gull+2009spec}.
  Previous work by \citet{bosl+2007} and \citet{koch+2007leo1} led to
  mean metallicities \feh\ $=-1.34$ and \feh\ $=-1.31$, respectively.
  Our photometric age-corrected metallicity therefore agrees with all
  recent spectroscopic measures.

\subsection{Comparison with spectroscopy}

\realfigure{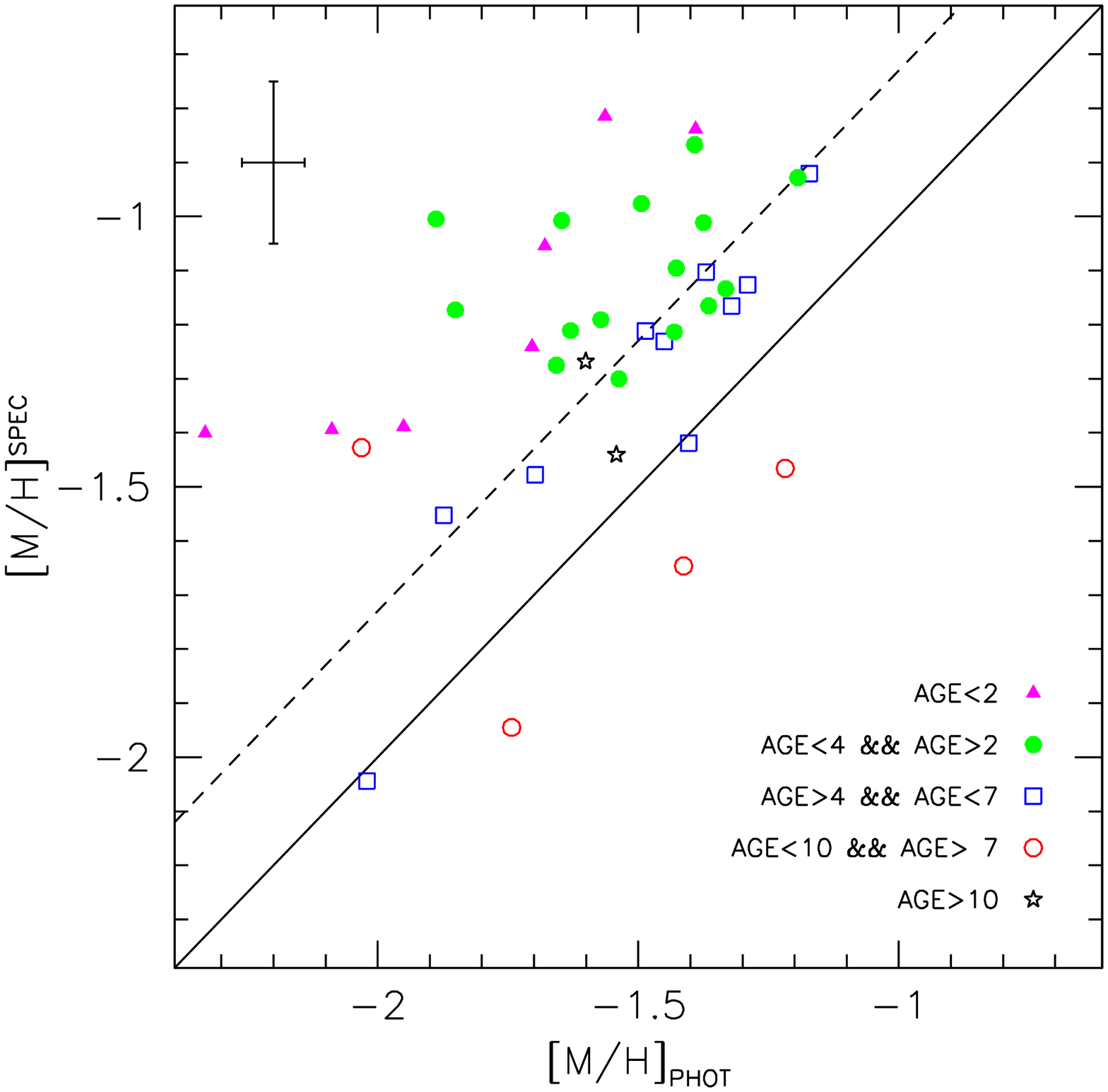}{
  Individual metallicity measures as derived from spectroscopy and
  from photometry, without any age correction.  Different symbols are
  used for stars with different ages, as derived by
  \citet{gull+2009spec}.  The {\it solid line} is the bisector, while
  the {\it dashed line} corresponds to the age correction for 5 Gyr
  old stars. The error bars represent the mean error of the
  spectroscopic metallicities and the r.m.s. (internal) error of [M/H] 
  from optical--NIR colours.
}{f:compspec2}

  The metallicities of individual RGB stars in \leoi, derived from
  NIR photometry, can be directly compared with the results of
  spectroscopic measurements.
  In Fig.~\ref{f:compspec1}, the metallicity distribution of \leoi\
  RGB stars inferred from \vks\ colours is compared with the
  spectroscopic metallicity distribution from \citet{gull+2009spec}
  (we refer the reader to that paper for a full account of previous
  spectroscopic work).  The photometric MDF was approximately 
  corrected for age population effects, assuming a common age of 5 Gyr
  for \leoi\ stars.
  The [M/H] distributions are similar, both in the
  mean values and the range.  The primary difference is in the
  low-metallicity tail of the photometric MDF.  This can be partially
  explained by a fraction of contaminating AGB stars, although the
  main reason is probably the presence of young stars with colour 
  bluer than the RGB, which are incorrectly treated as metal-poor
  stars.

\realfiguretwofig{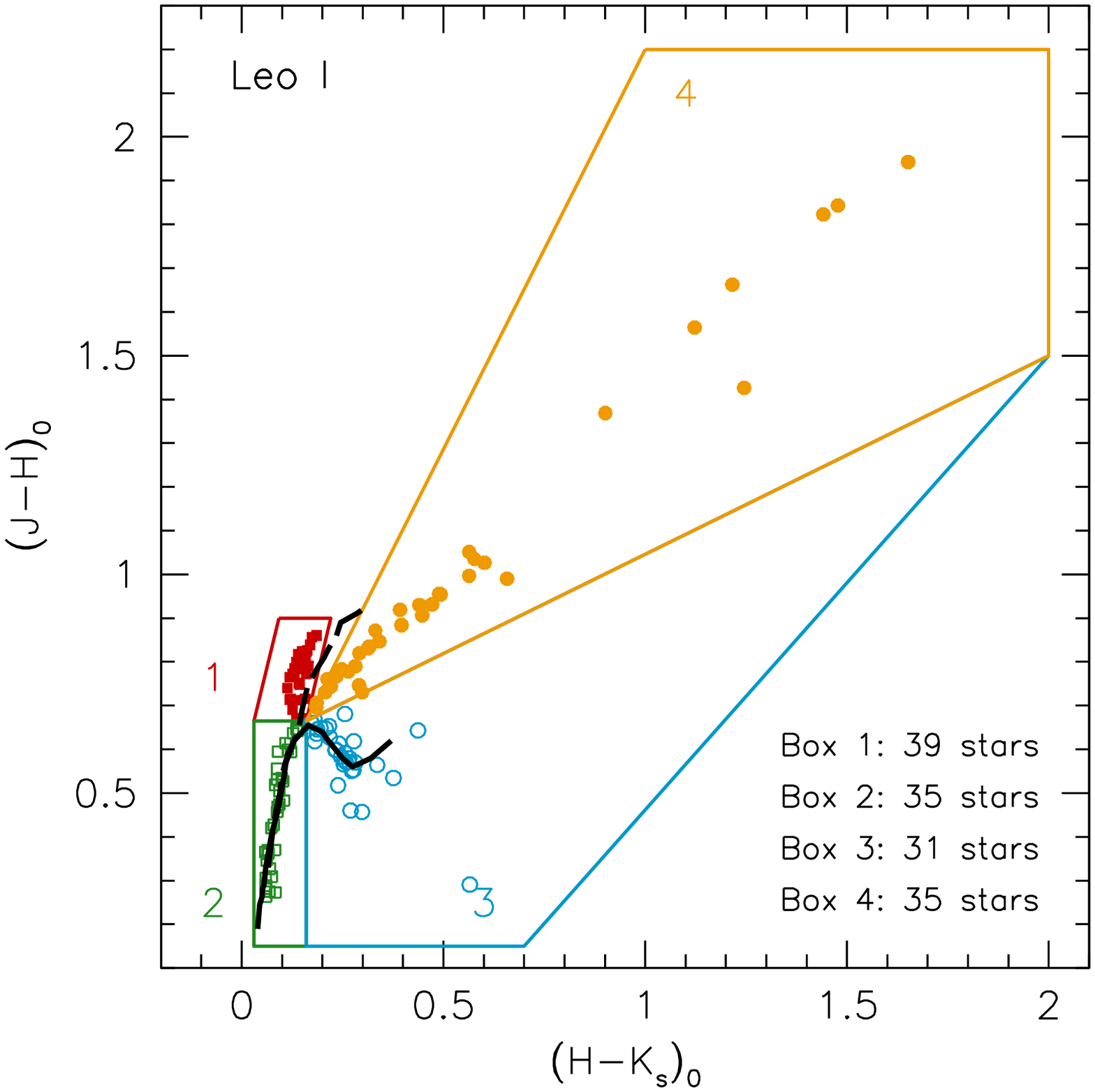}{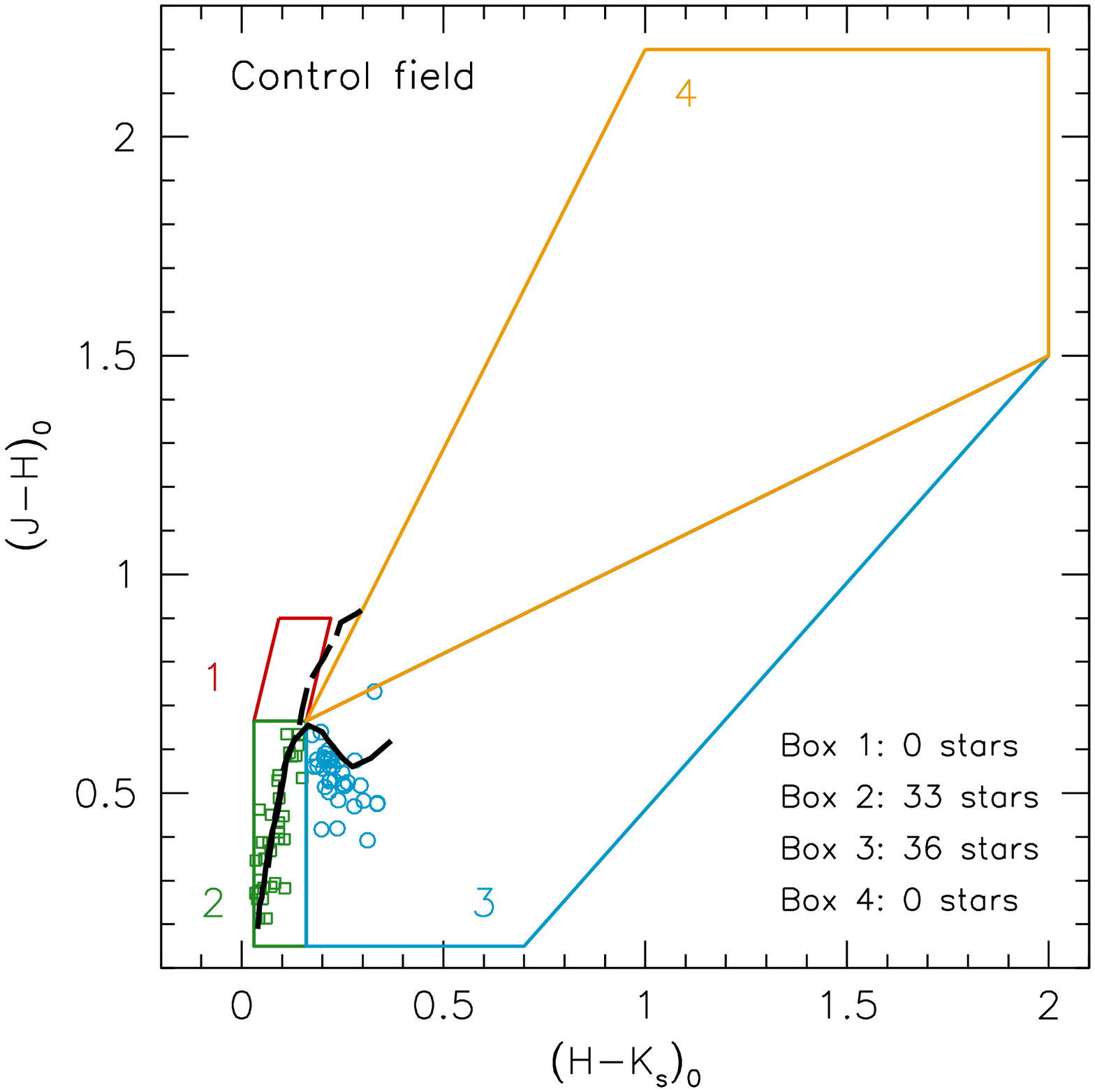} { 
  Near-IR two-colour diagrams of stars {\it brighter than the RGB tip} in
  the \leoi\ field ({\it left panel}) and the control field ({\it
    right panel}), 
  with superimposed the regions used to distinguish stars in \leoi\
  from foreground stars.  Region (1) delimits probable O-rich AGB
  stars in \leoi, regions (2) and (3) contain Galactic dwarf stars,
  region (4) is populated by C stars in \leoi.  Different symbols and
  colours are used for stars in different regions.
  Also shown are the loci of giant stars ({\it dashed line}) and
  main-sequence stars ({\it solid line}) from \citet{bessbret1988}.
  Note the absence of stars belonging to \leoi\ in the control field,
  which is located beyond the ``tidal radius'' of \leoi.
}{f:2colbox}

  The significant overlap between our sample of RGB stars and the
  dataset of \citet{gull+2009spec} (39 stars in common) also allows 
  a direct comparison of photometric and spectroscopic metallicity
  estimates on a star-by-star basis.
  Individual photometric and spectroscopic measures are plotted in
  Fig.~\ref{f:compspec2}, where stars in different age intervals
  \citep[as estimated by ][]{gull+2009spec}, are marked with different
  symbols. In this plot, the metallicities derived from photometry
  have not been corrected for the age difference.  
  Were all \leoi\ stars as old as the GGCs, we would expect the
  spectroscopic and photometric metallicities to be equal within the
  errors, and cluster about the bisector in Fig.~\ref{f:compspec2}.
  This is clearly not the case, as most stars are on the left of the
  unity relation, i.e. their metallicities are underestimated by the
  \vks\ colours. This is consistent with the mean age of the bulk 
  of the \leoi\ stellar populations.
  We have plotted an age-corrected relation (dashed line in
  Fig.~\ref{f:compspec2}), by calculating the metallicity correction
  to be applied to our photometric measures for young stellar
  populations ($\Delta$\mh~$=0.27$ dex, as above).  Indeed, many stars
  are close to this line, with the youngest stars located even more to
  the left, as expected.  The observed shift in \vks\ of young stars
  is consistent with our age determination based on optical colours.
  The metallicities derived using the \vks\ colours agree with
  the spectroscopic estimates, once stars' ages are accounted
  for. Alternatively, the \vks\ colours could be used to estimate the
  ages of stars once their metallicities are known from spectroscopy.

\section{AGB stars}
\label{s:leo1-AGB}

We now turn our attention to the AGB star content of \leoi.  Our WFCAM
$JH \kk$ catalogue provides the database for a study of the
intermediate-age AGB stars in \leoi\ with unprecedented photometric
accuracy and spatial coverage. From this database, we now select new
samples of C-rich and O-rich AGB stars and discuss their properties.

\subsection{Two-colour diagrams}

\realfigure{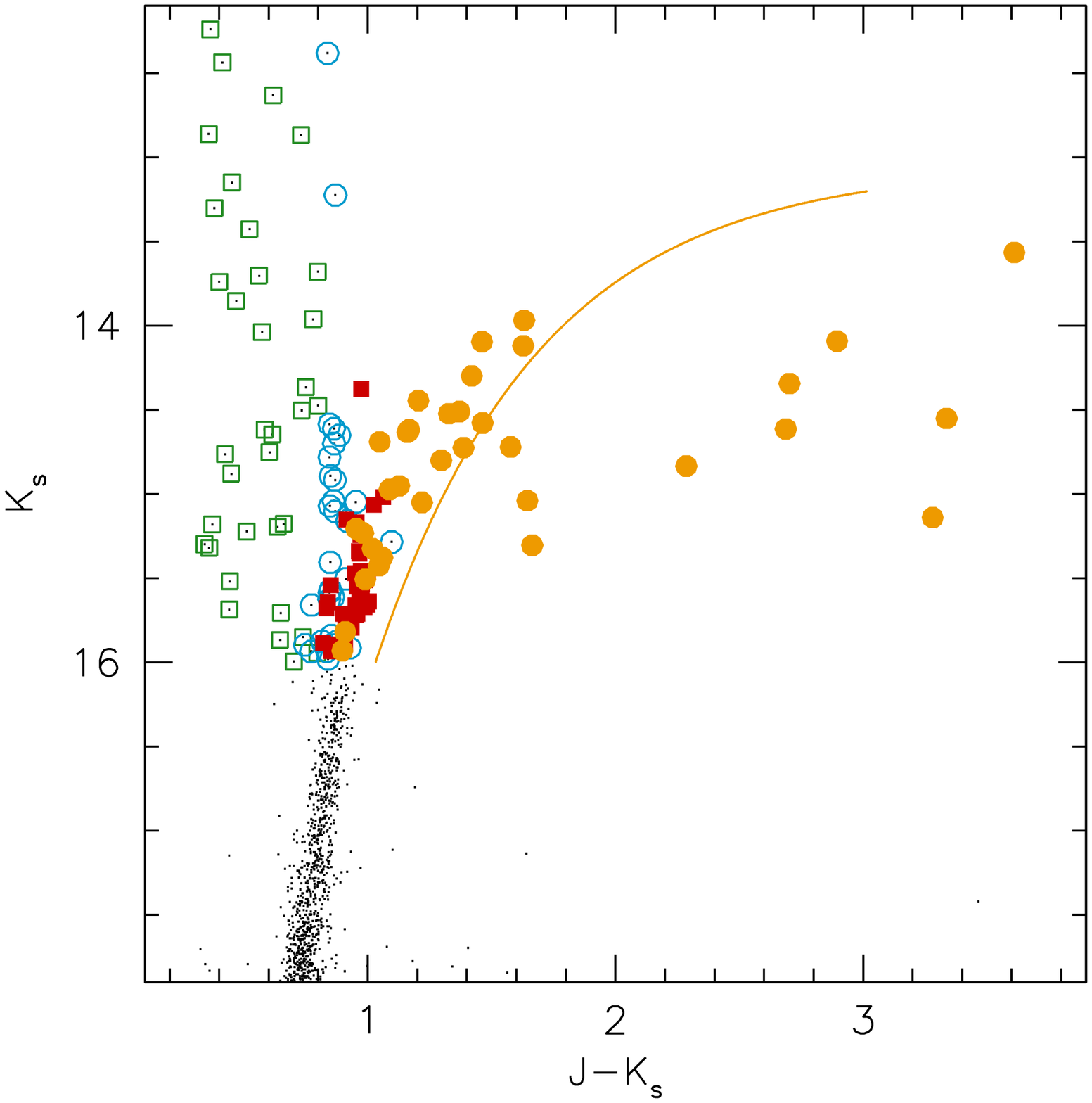}{
  The position of AGB stars and foreground stars, classified according
  to the NIR two-colour diagram of Fig.~\ref{f:2colbox}, in the
  CMD of the \leoi\ field.
  O-rich AGB stars (region~1, {\it red squares}) and C stars
  (region~4, {\it yellow filled circles}) in \leoi\ are effectively
  discerned from foreground Milky Way stars (region~2, {\it green open
    squares} and region~3, {\it cyan open circles}).
  The {\it solid curve} is a mean colour-magnitude relation for C
  star in Local Group dwarf galaxies \citep{tott+2000}, scaled to the
  distance of \leoi.
}{f:cmdbox}

  The two-colour diagram is an efficient tool to separate dwarf and
  giant stars \citep{aaromoul1985,bessbret1988}.  As such, it
  is particularly useful when studying extragalactic systems for 
  separating the stellar populations of the system from the
  Galactic foreground.

  Figure~\ref{f:2colbox} shows the NIR two-colour diagrams of stars
  brighter than $\kk =16.0$ in the field of \leoi\  and in
  a comparison field beyond the ``tidal radius'' of the galaxy.  The 
  magnitude cutoff was chosen to include only stars brighter than the
  RGB tip, leaving out first-ascent red giant stars.
  Following the approach of our \leoii\ study \citep{gull+2008leo2IR},
  we defined 4 regions in the two-colour diagram aimed at
  discriminating C stars from O-rich AGB stars in \leoi, both against
  the Galactic foreground. The star counts in the different regions
  are given in Fig.~\ref{f:2colbox}.

  Regions 2 and 3 contain the loci of giant and main-sequence Galactic
  stars from \citet{bessbret1988}.  The star counts are equal in
  \leoi\ and the control field within the statistical $\sqrt{N}$
  fluctuations, indicating a population of foreground
  Milky Way stars.
  Regions 1 and 4 are populated only in the case of the \leoi\ field,
  which proves that stars in those regions belong to
  the dSph galaxy.  The stars in region~1, which are close to the
  \citet{bessbret1988} line for giant stars, are identified with
  O-rich AGB stars in \leoi. 
  The photometry of candidate O-rich AGB stars in region~1 is given in
  Table~\ref{t:box1}. 
  Region~4 contains stars which become increasingly reddened in both
  \jh\ and \hks\ colours.
  In dSph galaxies this region is most likely populated by C stars
  \citep{gull+2008leo2IR,menz+2008,whit+2009}, although in the Milky
  Way and the Magellanic Clouds it may be populated by other objects,
  such as M-type (O-rich) Mira variable stars and OH/IR stars
  \citep{bessbret1988,nikowein2000,menz+2002}.
  Mid-infrared spectroscopy with Spitzer Space Telescope of AGB stars
  in Local Group galaxies has shown that at low metallicity the dusty
  winds and mass-loss rates of C stars are similar to those of
  Galactic AGB stars, while mass loss is 1-2 orders of magnitude lower
  in O-rich AGB stars \citep[see][for a discussion]{laga+zijl2008}.
  For example, this is the case for Fornax dSph, which has a
  metallicity similar to that of \leoi\ \citep{mats+2007,laga+2008}.
  All previously identified C stars in \leoi\ are indeed found in
  region~4. NIR photometry for the candidate C stars selected from our
  two-colour diagram is given in Table~\ref{t:box4}, along with their
  cross-identification with previous surveys (unfortunately, no
  matching was possible with \citealt{aaromoul1985} because finding
  charts or coordinates are not provided). Most of them are previously
  identified C stars, while 6 are new candidates.  Note that star
  No.~11242 (ALW-17), which is marginally included in region~4, is not
  a C star according to \citet{azzo+1986}.
  A few very red stars in region~4 were not identified in optical
  bands.  For all stars with missing optical photometry, visual checks
  on our EMMI images ruled out possible mismatches and confirmed that
  the optical counterparts fall below the detection threshold. A
  special case is star No.\,8336 (also M02-D), which is very bright in
  the optical ($V=18.7$) compared to other stars in Table~\ref{t:box4}
  and has a blue optical colour $B-V=0.04$, inconsistent with that of
  C stars.  Inspection of both optical and NIR images confirmed its
  star-like shape.  This object is likely to represent a chance
  superposition or photometric blend \citep[see][ for a similar case
  in SagDIG]{gull+2007sag}.

  Figure~\ref{f:cmdbox} shows the location of \leoi\ stars selected
  using the two-colour diagram in the CMD.  The O-rich AGB star
  candidates from Table~\ref{t:box1} are distributed on a nearly
  vertical sequence above the TRGB, overlapping in magnitude and
  colour with the faintest C stars.  Note that most of the AGB stars
  within 1 mag above the TRGB are O-rich stars.
  The stars in Table~\ref{t:box4} closely follow the mean locus of C
  stars in LG galaxies \citep{tott+2000} up to \jks~$\sim 1.7$. Those
  with redder colours show a large scatter consistent with the known
  variability of upper-AGB carbon stars and the onset of circumstellar
  dust envelopes \citep{menz+2002,whit+feas2000}.

\realfigure{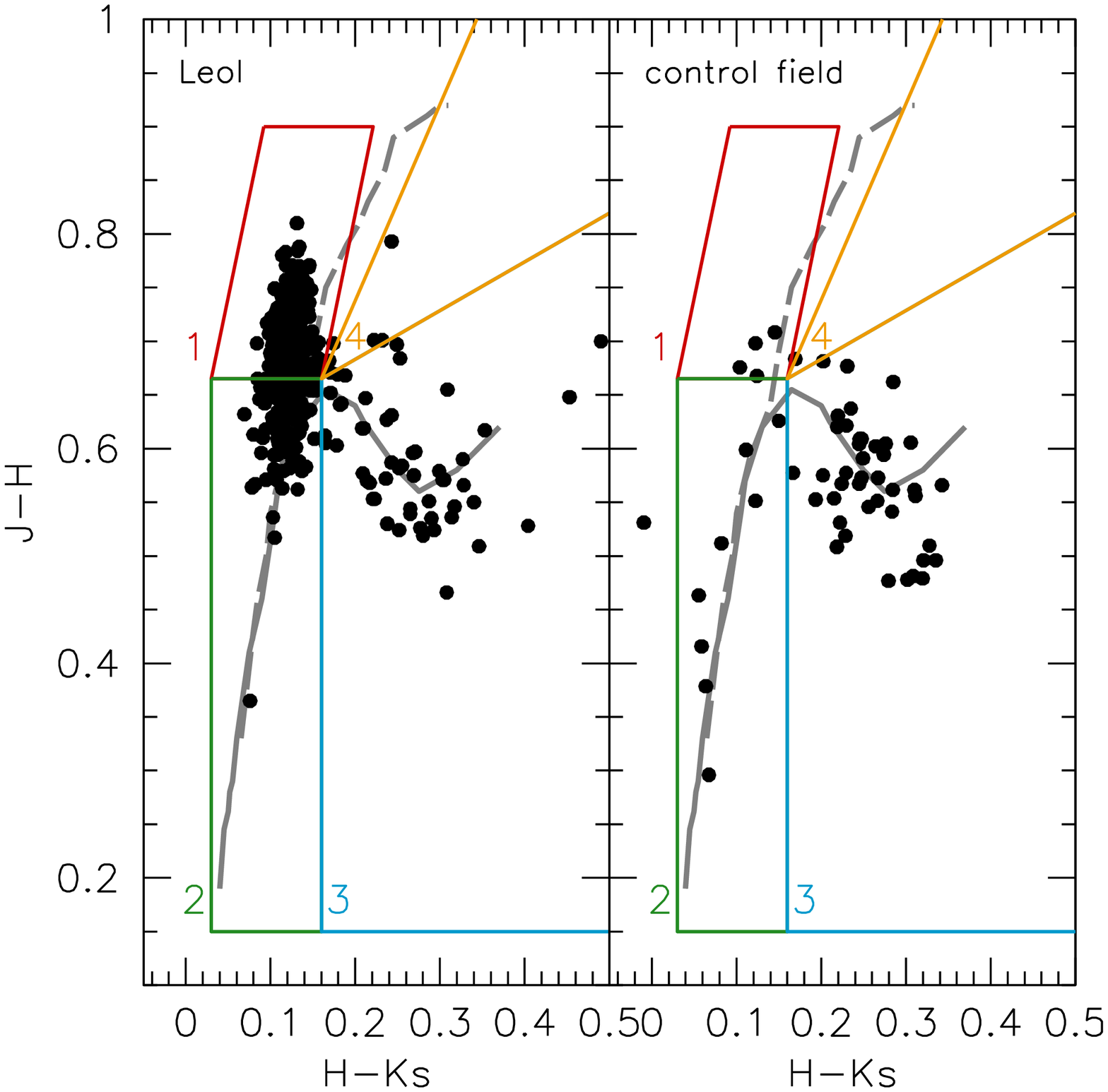}{
  Selection of a clean sample of RGB stars in \leoi\ using the NIR
  two-colour diagram.  In the {\it left} panel, we plot stars in the
  upper 1.3 mag of the RGB of \leoi\ dSph ($16.2 < \kk < 17.5$).  The
  same magnitude interval is shown in the {\it right panel} for the
  external field.
}{f:2colfield}

We note incidentally that the same two-colour technique can be used
for discriminating RGB stars in \leoi\ from the Milky Way foreground.
This is shown in Fig.~\ref{f:2colfield}, where we plot the NIR
two-colour diagram of stars in the magnitude interval $16.2 < \kk <
17.5$, which corresponds to the upper 1.3 mag of the RGB of \leoi.
The diagram is plotted for both the \leoi\ field and the control
field, and is slightly more scattered than the data in
Fig.~\ref{f:2colbox} because of the increased photometric error.
  The stars populating region~3 are mostly Galactic dwarfs.  This is
  confirmed by a $(V-\kk)$, $\kk$ plot of stars in the field of \leoi,
  where the wide baseline of the combined optical--NIR colour allows a
  neat separation of the foreground stars.

  The great majority of stars in regions~1 and~2 are red giants in
  \leoi. The near absence of stars in region~1 for the comparison
  field is easily explained by the fact that giant stars in the Milky
  Way are much brighter than the magnitude range of the RGB of \leoi.
  While the stars in region~1 are mostly in \leoi, the region~2 sample can
  be slightly contaminated by distant Galactic dwarf stars (but in the
  case of \leoi\ the contamination is negligible).  The two samples of
  RGB stars should probably be grouped together to avoid a bias in
  colour and metallicity.
  In conclusion, a selection based on broad-band $JH \kk$ imaging
  alone proves to be a valuable tool for selecting red giants stars.
  The increasing availability of wide-field NIR imagers makes this
  technique an interesting complement to intermediate-band optical
  searches \citep[e.g.,][]{maje+2000} to define clean samples of
  member stars in resolved systems for follow-up spectroscopic
  studies.

\realfigure{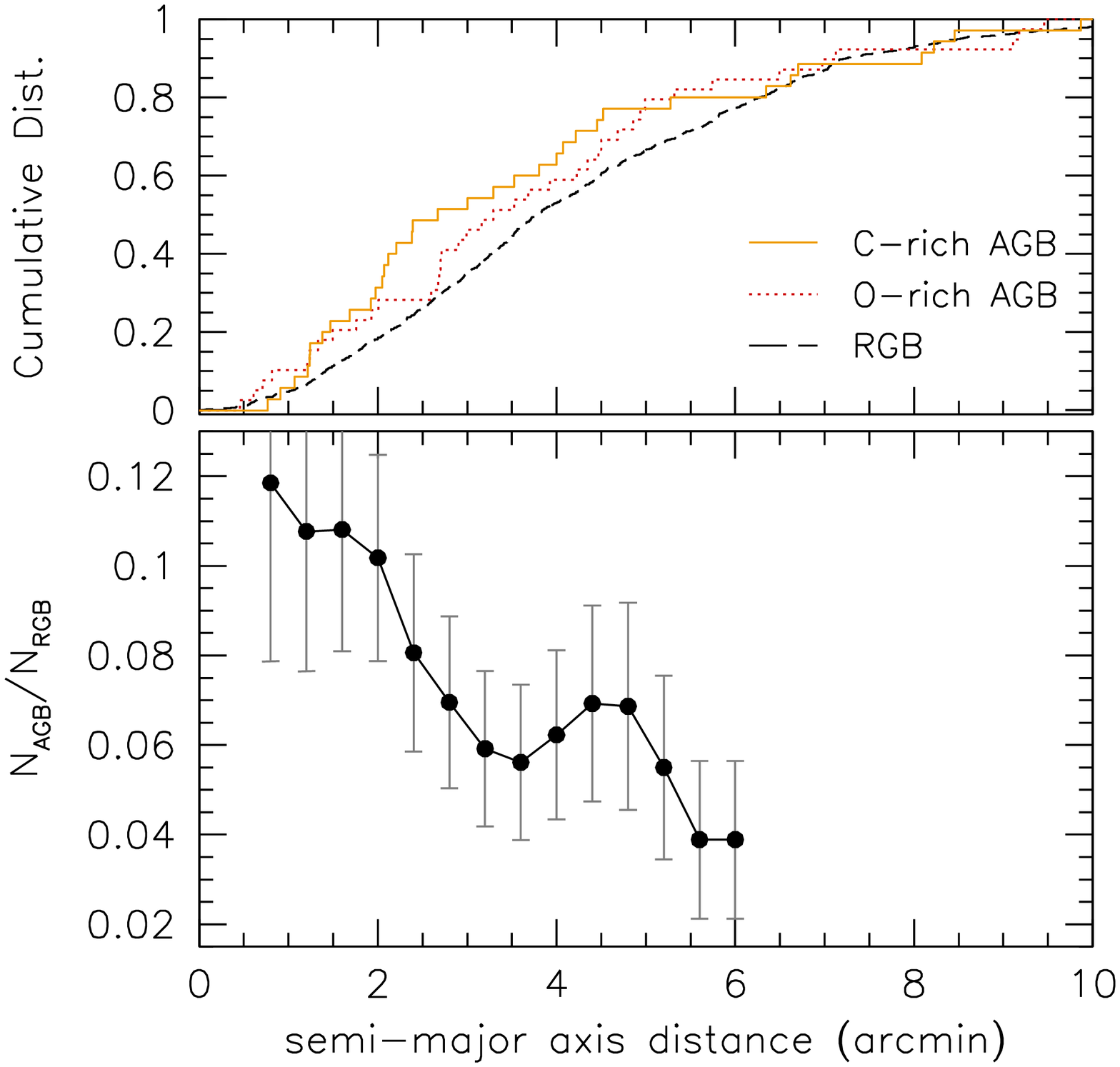}{
  {\it Upper panel:\ } comparison of the cumulative radial
  distributions of RGB stars and the samples of C- and O-rich AGB
  stars, as defined by our selection criteria based on the NIR
  two-colour diagram.
  {\it Lower panel:} radial variation in the number of AGB stars,
  normalised to the number of stars in the upper 2 mag of the
  RGB. Stars were counted in elliptical annuli with a 0\farcm4 step in
  semi-major axis (see text for details).  The error bars were
  obtained by error propagation of Poisson errors on stars counts.
}{f:distrad}

\subsection{Radial gradient in the AGB population}
\label{s:leo1radial}

\realfigure{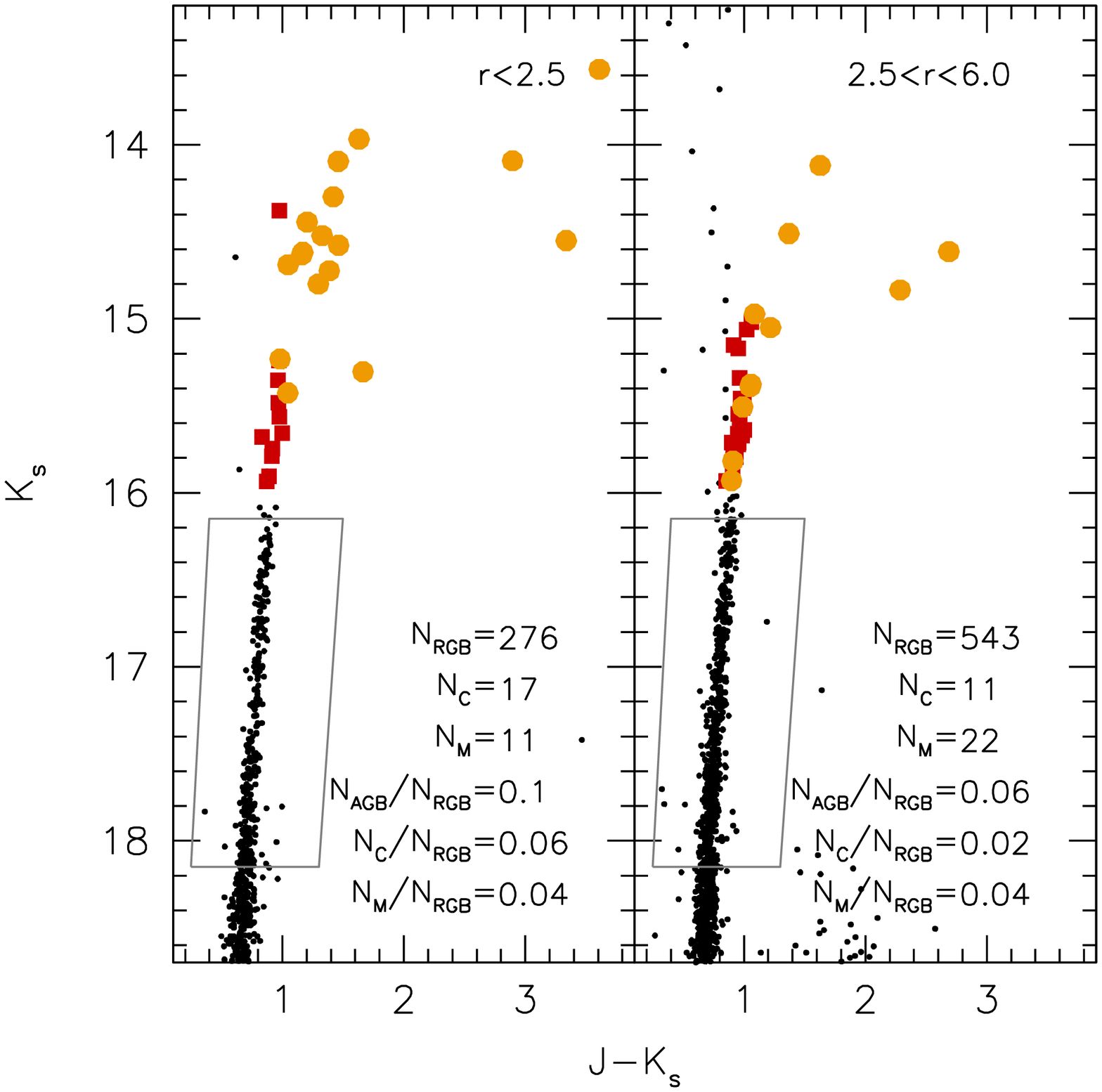}{
  Change in the AGB star population as a function of the 
  distance from the centre of \leoi.
  {\it Left:} CMD of stars in an inner elliptical region of \leoi\
  with semi-major axis $r = 2\farcm5$.  A box shows our selection of
  RGB stars.  O-rich AGB stars (indicated by the letter M, {\it
    squares}) and C stars ({\it circles}) were selected from the
  NIR two-colour diagram.
  {\it Right:} The same, except for an elliptical ring between $r=
  2\farcm5$ and $r= 6\farcm0$.  Outside the central region, the 
  \referee{fraction of C
  stars relative to the RGB stars is lower by a factor of 3}.
}{f:cmd_radial}

\leoi\ is one of the few dSph for which no evidence of a population
gradient was known from the literature.  Recently, our spectroscopic
study of \leoi\ \citep{gull+2009spec} gave a marginal detection of a
radial metallicity gradient among \leoi\ RGB stars, in contrast with
the conclusions of \citet{koch+2007leo1}.  We also found evidence for
a concentration of younger stars in the central region of \leoi.
  Then it is important to confirm the presence of stellar
  population gradients using AGB stars as tracers of intermediate-age
  stellar populations.  Radial distances were therefore calculated for
  AGB and RGB stars assuming that the centre of \leoi\ is located at
  $\alpha = 10^{\rm{h}}08^{\rm m}26 \fs 68$, $\delta = +12\degr 18
  \arcmin 19\farcs7 $ (J2000) and the stellar density
  distribution is approximately elliptical with ellipticity
  $\epsilon=0.37$ and position angle P.A.=$84\degr$ \citep{sohn+2007}.
  The radial distribution of AGB stars, normalised to the number of
  stars in the upper 2 mag of the RGB, is plotted in
  Fig.~\ref{f:distrad} (lower panel).  C-\  and O-rich AGB stars
  were counted in elliptical annuli with a 0\farcm4 step in semi-major
  axis.  The errors on the relative number of AGB stars were
  calculated assuming Poisson statistic for the star counts.
  A central concentration of AGB stars relative to RGB stars is
  apparent in the plot, with a decline of a factor of $\sim3$ at
  $6\arcmin$ with respect to the centre.
  The statistical significance of this result is larger when we
  consider C stars only (Fig.~\ref{f:distrad}, upper panel).  A
  Kolmogorov-Smirnov (\abbrev{KS}) test used to compare the cumulative
  distributions of AGB stars (C- and O-rich) to that of RGB stars gave
  the following results.  The null hypothesis that AGB and RGB stars
  are drawn from the same parent population can be rejected with a
  95\% confidence; the probability rises to 97\% if we consider only C
  stars. For the O-rich AGB stars, the null hypothesis cannot be
  rejected at any significant level.  These tests imply that the
  C-type AGB stars show a statistically significant central
  concentration, whereas the spatial distribution of O-rich AGB stars
  is similar to that of the bulk population of red giant stars.

  The central excess of C stars is also evident in
  Fig.~\ref{f:cmd_radial}, where we compare AGB stars in an inner
  $2\farcm5$ region of \leoi\ and in an elliptical ring with $2\farcm5
  < r< 6\farcm0$.  
  \referee{While the fraction of O-rich AGB stars relative to the
    number of RGB stars is constant in the two regions, the fraction
    of C-stars is 3 times higher in the central region}.
  Again, we conclude that the difference in the radial distribution of
  AGB and RGB stars is driven by an excess of C stars, which trace the
  intermediate-age stellar population in the age range 1--5 Gyr.  {\it
    This confirms the presence of an age gradient in \leoi\ as
    suggested by spectroscopy of RGB stars} \citep{gull+2009spec}.
  Indeed, the number of C-stars is much more dependent on age and SFH
  than the number of O-rich AGB and RGB stars, as illustrated in
  Fig.~\ref{fig_sfr} and discussed in more detail in the next section.
  Also, the mean age of red clump stars in \leoi\ is $\sim 5$ Gyr,
  definitely older than that of luminous AGB stars, which explains why
  no age gradient was detected by \citet{held+2000}.

\section{Comparison with theoretical models}
\label{s:leo1teo}

  The present observations sample quite completely the
  AGB population of \leoi\ in the surveyed area, except perhaps for
  the stars most absorbed by circumstellar dust. As discussed by
  \citet{gull+2008leo2IR} for \leoii, this kind of data is important
  for testing thermally-pulsing AGB (\abbrev{TP-AGB}) models in the
  regime of low metallicity. In the case of \leoi, the presence of
  recent star formation makes it a particularly interesting case for a
  study of AGB stars of different ages and masses.

  Since the HST observations used in the determination of the SFH of
  \leoi\ were obtained in a central region, and given the presence of
  a significant gradient in the stellar populations of \leoi, we
  performed our analysis on the inner (projected) 2.5~arcmin of the
  galaxy. Inside this area, the $N_{\rm AGB}/N_{\rm RGB}$ ratio is
  almost constant and comprised between 0.10 and 0.12.
  Selecting this small region also dramatically reduces the number of
  foreground stars, whereas the number of
  \leoi\ stars is reduced by just a factor of 2 relative to the total
  observed area (see Fig.~\ref{f:cmd_radial}).

\subsection{Simulating the photometry}

The \leoi\ star counts can be compared with the recent set of
  TP-AGB evolutionary tracks from \citet{marigira2007}.  The procedure
  followed to simulate the \leoi\ observations is essentially the same
  as in \cite{gull+2008leo2IR}, to which the reader is referred for
  details.  Suffice it to mention that the latest version of the
  \trilegal
\footnote{\tt http://stev.oapd.inaf.it/trilegal} 
population synthesis code \citep{gira+2005,giramari2007} was used to
simulate both the Milky Way foreground and the \leoi\ dSph.
For this simulation, we assumed for \leoi\ a distance of 250 Kpc 
and zero reddening in the near-infrared.  The simulations of
the TP-AGB include important effects such as luminosity and
temperature variations during thermal pulse cycles, and the processing
of radiation by circumstellar dust according to the \cite{groe2006}
prescriptions \citep[details are given in][]{mari+2008}. 

\realfigure{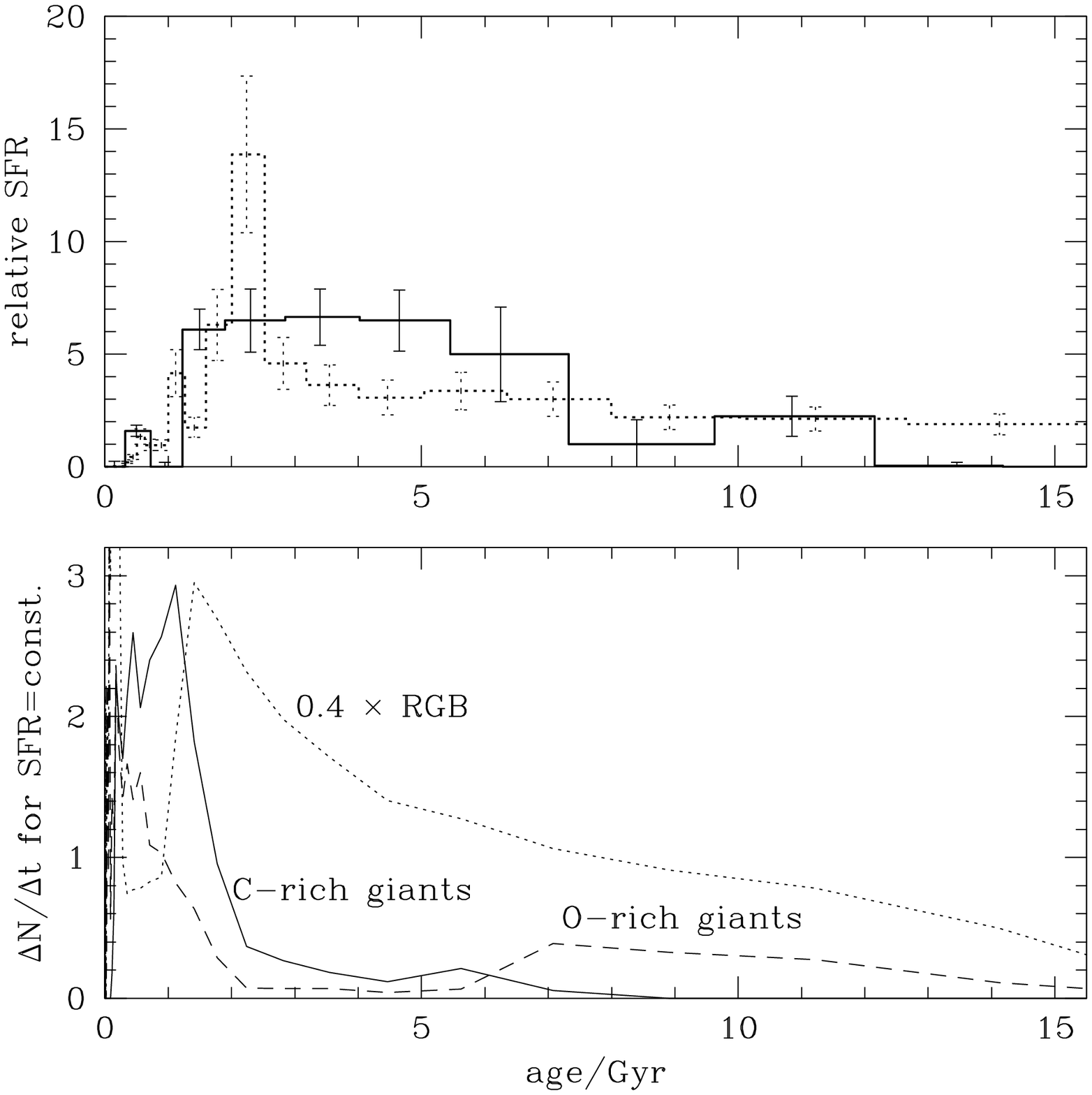}{ 
  {\it Top panel:} The SFHs used in this work: \citep[][dotted
  lines]{dolp+2005} and \citep[][solid lines]{gall+1999}.
  {\it Bottom panel:} The age distribution (number of stars per age
  interval) of different types of stars as a function of age, for a
  galaxy model forming stars at a constant rate from 0 to 15 Gyr, at a
  constant $Z=0.0005$ metallicity. The stellar kinds plotted are RGB
  stars within 2~mag of the TRGB (multiplied by 0.4; dotted line), and
  both O-rich (dashed line) and C-rich (solid line) giants within
  2~mag above the oldest TRGB.}{fig_sfr}

The total mass of simulated \leoi\ stars inside the area of our
  observations was chosen so as to reproduce the star counts in the
  upper 2 magnitudes of the RGB, for which our observations are
  complete. 
  The relative star-formation rate of \leoi\ was taken from two
  different sources, as depicted in the upper panel of
  Fig.~\ref{fig_sfr}: \citet[their figure 9]{gall+1999} and
  \citet{dolp+2005}. In both cases the SFH was derived by inversion of
  a deep CMD from HST observations.  Both determinations indicate that
  the star-formation rate of \leoi\ increased in the last few Gyr.
  This feature appears as a broad peak in the SFH between $\sim 1$ and
  7 Gyr in the case of \citet{gall+1999}, and as a much more confined
  peak, close to 2.5 Gyr, in the case of \citet{dolp+2005}.
  For the \leoi\ metallicity, 
  we adopted a Gaussian distribution with mean
  [M/H]$=-1.3$ and dispersion 0.2~dex. This metallicity distribution
  was assumed to be the same for all ages. This is a fair 
  approximation
  since the bulk of the AGB stars arise from an age
  interval where spectroscopy suggests modest chemical evolution
  \citep{gull+2009spec}.  

  To reduce the statistic fluctuations in the numbers of predicted
  stars, each simulation was run 100 times with different random
  seeds. One example of such simulations is shown in
  Fig.~\ref{fig_simcmd}, which reproduces quite well the observed
  diagram shown in Fig.~\ref{f:4cmd}. The main stellar features are
  accounted for by the model, including the colour distribution of C
  stars which, with the new prescriptions for circumstellar dust
  extinction, turns out to be consistent with the extended AGB red
  tail in the data.
  The simulations do not contain the objects found in the bottom-right
  part of the diagram ($K>18$, $J-K>1.2$), which likely correspond to
  background galaxies \citep[see, e.g., ][]{nikowein2000}.

\realfigure{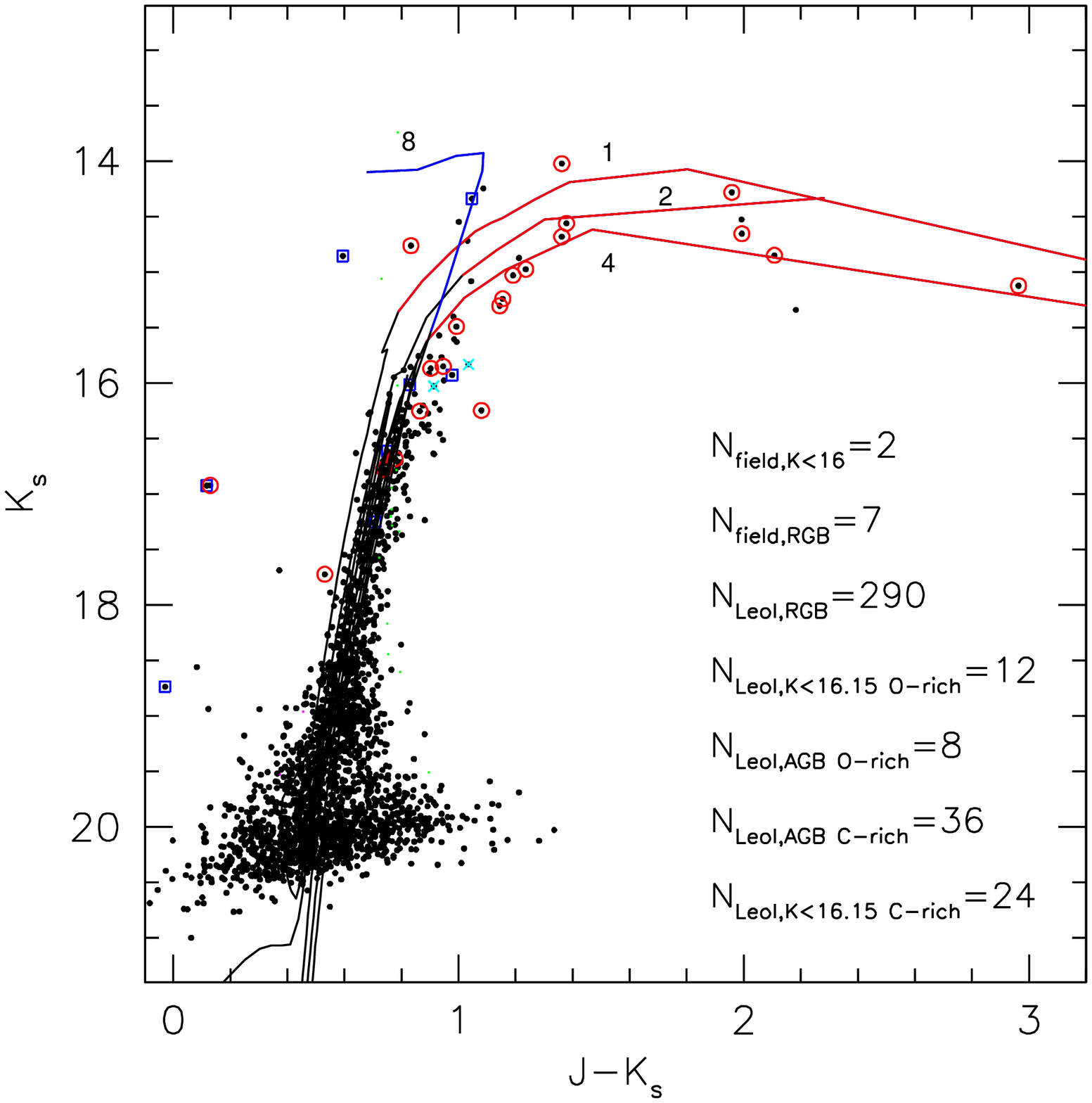}{
  One of the many simulated CMDs for \leoi, using the
  \citet{dolp+2005} SFH. 
  In the electronic version of this paper, different
  colours mark different kinds of stars, namely: Milky Way disk
  (green) and halo (magenta), \leoi\ pre-TP-AGB stars (dark), and
  \leoi\ TP-AGB stars both O-rich (blue) and C-rich
  (red).}{fig_simcmd}

\subsection{Comparing foreground and RGB counts}

  According to the model counts, the expected number of foreground
  Milky Way stars in our $3.4\times10^{-3}$~deg$^2$ area (i.e. the
  area of an ellipse with semi-major axis $r = 2\farcm5$ and
  ellipticity $0.37$), in the $13< \kk <16$ magnitude interval, is
  $2.9 \pm 1.7$; this agrees with the 3 objects observed in regions 2+3
  at $\kk < 16$ and $r < 2\farcm5$.
  In view of this agreement, our simulations can be used to infer the
  field contamination in other CMD regions. We find that a total of
  $6.5\pm2.3$ foreground stars are expected to contaminate the
  uppermost 2~mag of the RGB in \leoi, in the $16.2< \kk <18.2$
  interval.
The observed number of stars in this CMD region is 276
  (Fig.~\ref{f:cmd_radial}), which largely outnumbers the expected
  foreground stars. We thus conclude that about 270 genuine RGB stars
  in \leoi\ are expected within 2 mag of the tip. The \leoi\
  simulations are then scaled so as to obtain a total of $270\pm18$
  RGB stars in the upper 2 magnitudes of the RGB.

\subsection{Comparing AGB counts}

Using 100 simulations for each SFH, we obtained the following
  results for AGB star counts in \leoi.
For the \citet{dolp+2005} SFH, we expect to find
$14.8\pm4.1$
O-rich giants above the RGB tip. Among them, 
$11.6\pm3.7$ 
are genuine TP-AGB stars, the remaining being either early-AGB or core
helium burning stars of the youngest ages. The predicted C-rich AGB stars are
$40.6\pm6.9$ 
in total, 
$30.2\pm5.9$ 
brighter than the TRGB in the $K$ band.
In the inner $2\farcm5$, the data present $11$ O-rich and $17$ C-rich
stars in regions 1 and 4 of Fig.~\ref{f:2colbox}.  While the number of
O-giants in the simulations agrees with the observed one, the excess
of C-rich giants is significant, by a factor of about 2.
Alternatively, using the \citet{gall+1999} SFH, we predict
$10.5\pm3.4$
O-rich giants above the RGB tip 
($7.0\pm2.8$ 
genuine O-rich TP-AGB), and 
$47.2\pm6.9$ 
carbon stars 
($34.0\pm5.7$ 
above the TRGB). These numbers indicate a small deficit of O-rich
stars (not significant, however), and again an excess in the 
predicted number of C stars by a factor of $\sim 2$.

\realfigure{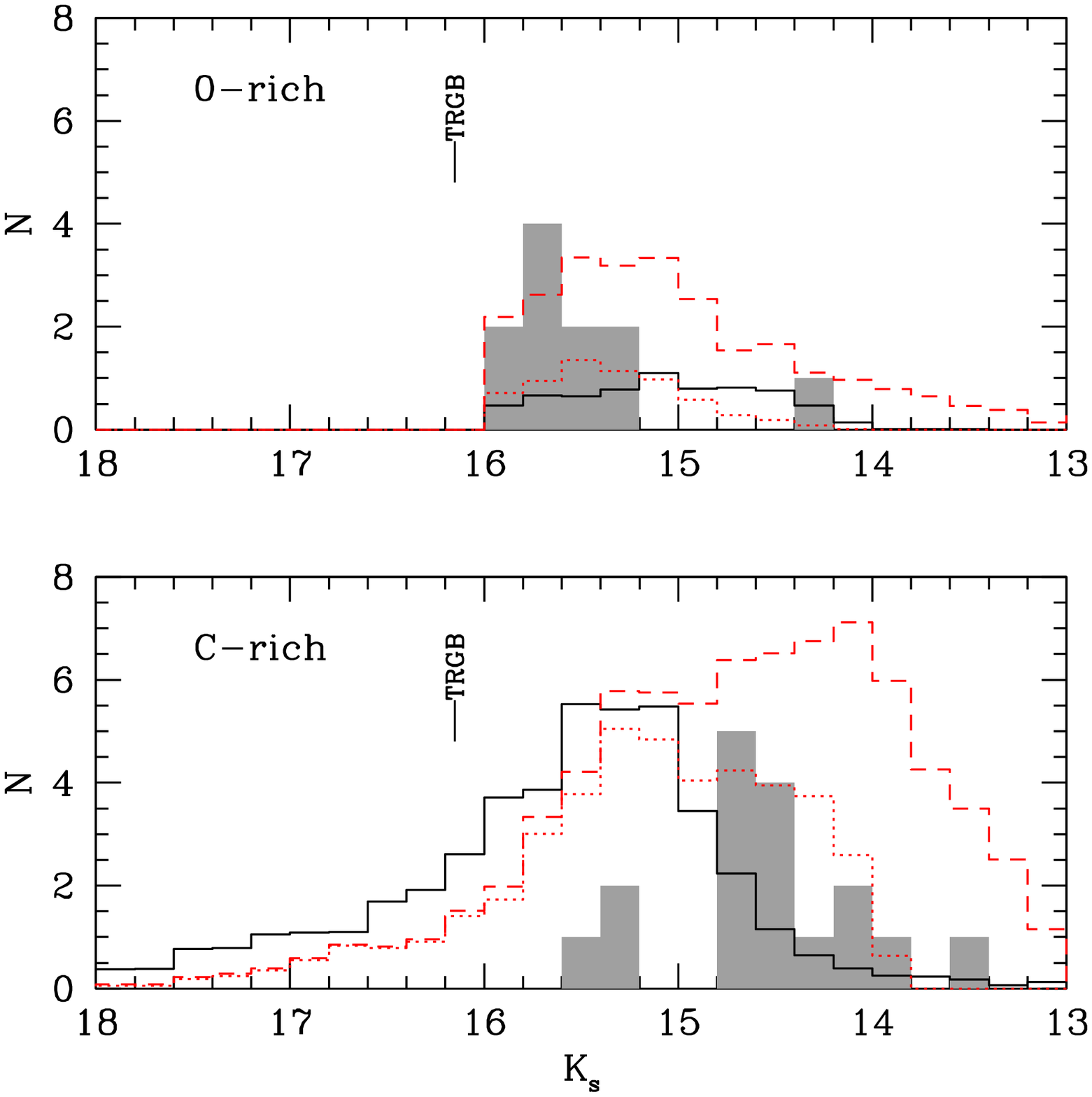}{
  Simulated LFs for luminous stars in \leoi\ ({\it solid line}),
  separated as O-rich giants above the TRGB (top panel) and C-rich
  TP-AGB stars (bottom panel), for the \citet{gall+1999} SFH. The {\it
    shaded histograms} correspond to all candidate AGB stars in
  Tables~ \ref{t:box1} and \ref{t:box4}. Also shown for comparison is
  the simulated LF for the SMC, both including ({\it dashed line}) and
  excluding ({\it dotted line}) stars younger than 1.5 Gyr.
}{fig_simlf}

The mean C-star luminosity function
is compared to the data in Fig.~\ref{fig_simlf} (bottom panel).  
As suggested by a comparison between the data in Fig.~\ref{f:4cmd}
and the simulations in Fig.~\ref{fig_simcmd}, the simulated C stars
are fainter on the mean than the observed stars.  A KS test indicates
a negligible probability that the observed distribution is drawn from
the model distribution.
\referee{In particular, we note that the predicted number of C stars
  fainter than the RGB tip, which is significant in our models,
  contrasts with the paucity of faint C stars in the spectroscopic
  surveys (although the completeness of the observational searches of
  C stars for objects fainter than the RGB tip is not well
  understood).}

  A similar comparison for O-rich AGB stars above the RGB tip
  (Fig.~\ref{fig_simlf}, top panel) shows that the models predict a
  distribution extended to brighter magnitudes than observed, with a
  KS probability that the observed distribution is drawn from the
  theoretical model of just 1.5\%.  Although the number of O-rich AGB
  stars seems to be correctly predicted, the stars are brighter than
  observed by $\sim 0.5$~mag.

  \referee{ Thus, as in the case of \leoii, there are significant
    discrepancies between the LFs predicted by current low-metallicity
    TP-AGB models and the observed data.  However, the trends are
    quite different in the two cases.  In \leoii\ the lifetimes of
    O-rich AGB stars are overestimated, while the C stars are
    reproduced in seemingly acceptable numbers
    \citep{gull+2008leo2IR}.
  For \leoi, the situation regarding O-rich and C-rich AGB star counts
  is the opposite.  The difference is likely related to the different
  mean ages of AGB stars in these two galaxies, since the \leoii\
  stellar populations are significantly older on average than those in
  \leoi.
  The few observed C stars in \leoii\ sample the lower limit of the mass
  interval for which C stars can be formed by thermal pulses, with
  masses close to 1.0 \Msun\ and ages between $\sim 5$ and 7~Gyr.
  Conversely, the numerous C stars in \leoi\ sample the $M>1.3$ \Msun\
  interval for ages younger than 3 Gyr ($Z=0.001$). The differences
  found for C stars are therefore not surprising.}

\subsection{Comparison with the SMC}

The current models of TP-AGB stars were calibrated using
  constraints from the Large and Small Magellanic Clouds, in
  particular their global C-star luminosity functions, and AGB
  lifetimes derived from their star clusters \citep{marigira2007}.
  Therefore, the behaviour of these models reflects constraints set by
  stellar populations with metallicity between \feh~$= -1.0$ and
  $-0.4$.
  \leoi\ has a mean metallicity $\sim -1.3$ dex, which is 0.5 
  dex lower than that of the young SMC populations, and comparable to
  that of intermediate-age and old populations in the SMC.
Thus, it is not clear why models calibrated on the SMC observations
predict the numbers and luminosities of the \leoi\ AGB stars so
poorly.  Is the moderate difference in metallicity the main reason of
the observed discrepancies ?

To clarify these questions, we made two tests.
First, we tested the effects of changing the metallicity on the
LF. Simulations were produced with metallicity varying from [M/H]~$=
-1.6$ to $-1.0$.  The resulting LFs were nearly the same as those
obtained for [M/H]~$=-1.3$, indicating that the problem cannot reside
in the metallicities.
Second, we compared the properties of AGB stars in
\leoi\ and the SMC.  New SMC simulations were performed using the SFH 
of \citet{harrzari2004} together with the age-metallicity relation of
\citet{pagetaut1998}.  We used the \trilegal\ code in its current 
version, yielding more detailed simulations than the original code 
presented in \citet{marigira2007}.
The new simulations, which have the same features (e.g., detailed
modelling of the photometry) and the same physics as used for \leoi,
essentially confirm the results of \citet{marigira2007} concerning the
shape of the C-star LF and the good overall fit to the SMC data.  This
SMC model is shown in Fig.~\ref{fig_simlf} (dashed histogram),
shifted to the distance and reddening of \leoi\ dSph.
The C stars in the SMC are apparently much brighter on the mean than
those in \leoi\ (continuous histogram, using the Gallart et al. SFH).
The dotted histogram shows the same model for the SMC, but 
excluding the stars with ages younger than 1.5 Gyr.
It is then clear that the bulk of the bright C stars in the SMC comes
from the strong star formation episode experienced in the recent past
\citep[see ][]{harrzari2004}.

\realfigure{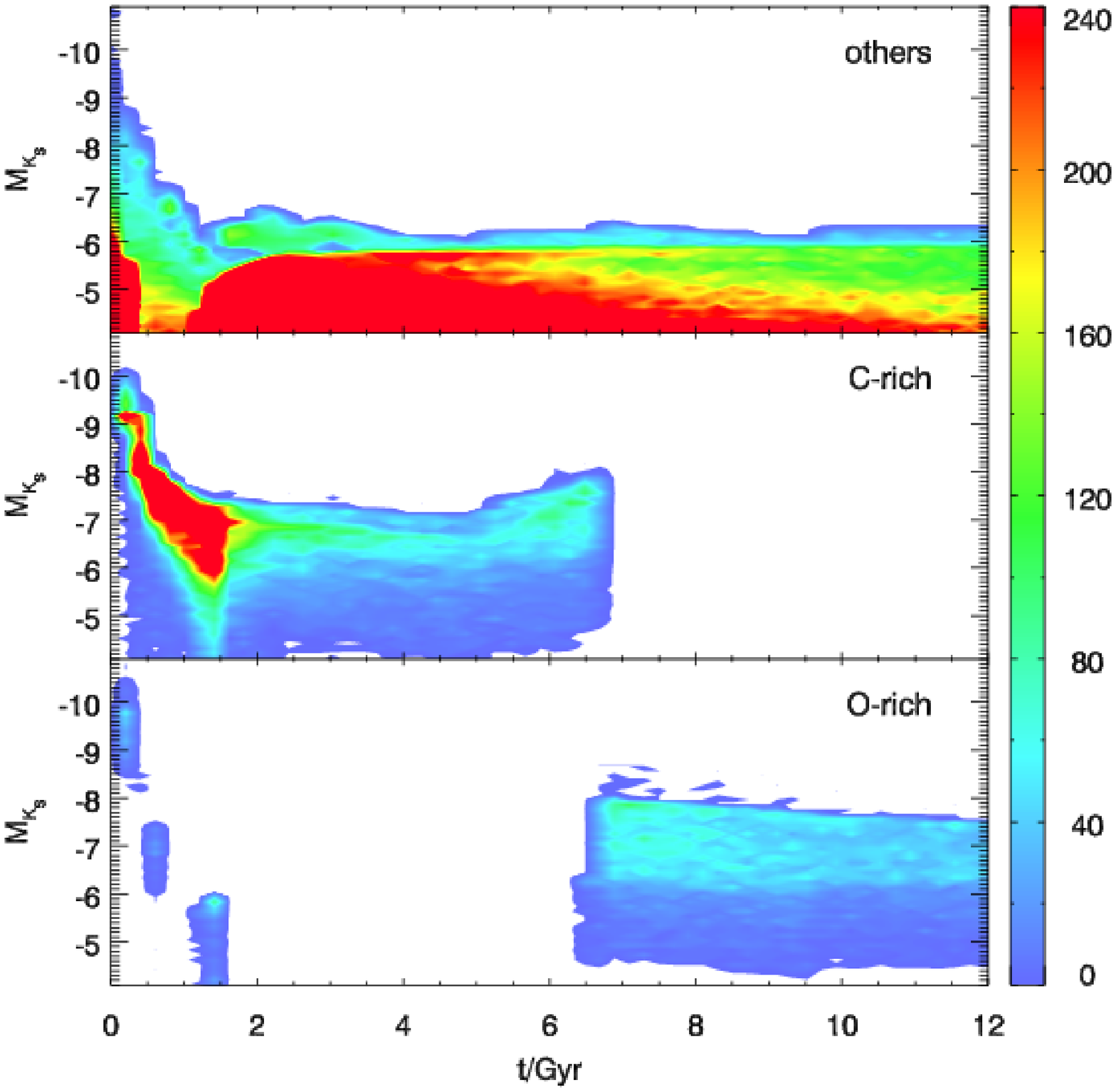}{
  Simulated number density of evolved (RGB and AGB) star in a stellar
  system with constant star-formation rate and metallicity [M/H]~$=
  -1.3$.  The number of stars is colour-coded as a function of age and
  $\kk$-band absolute magnitude.
  {\it Top panel:\ } stars that did not undergo any thermal
  pulses. These stars comprise RGB stars and early-AGB stars (older
  than $\sim 1$ Gyr) and red supergiants evolving from
  intermediate-mass stars younger than 1 Gyr.
  {\it Middle panel:\ } luminosity distribution of C-rich AGB stars.
  {\it Bottom panel:\ } O-rich TP-AGB stars.  
}{fig_map_agb_models}

The C-star LF is clearly a sensitive function of the SFH of the
system. This is confirmed by Fig.~\ref{fig_map_agb_models}, where we
present a colour-coded plot of the number of AGB stars in \Mks\ and
age bins in the case of a constant star-formation rate and fixed
metallicity equal to that of \leoi.
In Fig.~\ref{fig_map_agb_models}, most of the TP-AGB stars younger
than $\sim 6.5$ Gyr become C stars soon after they enter the TP-AGB
phase, except for some TP-AGB stars in the 2--5 \Msun\ range which
also appear as O-rich because either are evolving through the first
part of their TP-AGB phase before being converted to C stars, or
(those younger than 0.5 Gyr) are massive enough to experience
hot-bottom burning, thus keeping their surface C/O ratio $\la 1$
\citep[e.g., ][]{mari2007}.
As seen in the middle panel of Fig.~\ref{fig_map_agb_models}, a
sustained star formation activity in the last 1--2 Gyr does imply a
high production rate of C stars.  Since a recent episode of star
formation ($\le 1$ Gyr) is absent in \leoi\ dSph, very luminous C
stars are missing in the simulations. 

While the C stars in the SMC mostly sample the properties of young C
stars of metallicity close to $-0.4$, with initial masses higher than
$\sim 2$ \Msun, the AGB stars in \leoi\ mainly arise from
intermediate-age stars with metallicity $-1.3$ and initial masses
close to 1.5 \Msun.

\subsection{Models vs. observations}

Then, for the bright end of the C-star LF, it is conceivable that the
star-formation histories used to calculate the model LF are biased
towards stellar populations which are on average too old.
  However, it appears likely that the models actually overestimate the
  lifetime of low- and intermediate-mass C-type stars, and/or
  underestimate their luminosities.  There are different ways by which
  the lifetimes/luminosities may be changed: by increasing the
  mass-loss rates for C stars, by decreasing the efficiency of
  dredge-up events (so as to more rapidly increase the core mass and
  luminosity), or by efficient dimming of their near-infrared light by
  circumstellar dust shells.
    Another aspect that might be potentially relevant to this issue is
    the fact that at the low metallicities considered here the TP-AGB
    models attain a surface C/O~$>1$ already during the first few
    sub-luminous thermal pulses, before reaching the core-mass
    luminosity relation characteristic of the fully-developed TP-AGB
    regime \citep[][]{wage+groe1998}.  Though the effect of a very
    efficient third dredge-up on the stellar luminosity during these
    earlier stages is still not well assessed, we find in the
    literature some theoretical indication that TP-AGB stars might be
    brighter than commonly assumed by synthetic models \citep[see,
    e.g., ][]{herw+1998,mowl1999}.
    While exploring the possible solutions requires the calculation of
    new grids of TP-AGB evolutionary models, which is beyond the scope
    of this paper, the present results \citep[along with those
    of][]{gull+2008leo2IR} demonstrate that, despite the possible
    uncertainties in SFHs and age-metallicity relations in nearby
    dwarf galaxies, the observed properties of AGB stars in dSph
    galaxies provide valid constraints to AGB models at low
    metallicities.

\section{Summary and conclusions}
\label{s:leo1summ}

  We have presented new infrared $JH \kk$ photometry of \leoi\ dSph
  obtained with the wide-field imager WFCAM at the UKIRT.  Our NIR
  photometry, combined with optical $B$ and $V$ band photometry, 
  provides information on the spectral energy
  distribution of AGB and RGB stars in \leoi\ on a wide colour
  baseline.

  We provide a catalogue of $JH\kk$ PSF-fitting photometry of evolved
  stars in \leoi\ down to $\kk \sim 20.5$, over a $13\farcm5 \times
  13\farcm5$ area, from which the basic parameters of \leoi, distance
  and metallicity, were re-derived. We measured the distance to \leoi\
  from the $J$, $H$ and $\kk$ magnitude of the TRGB, using a
  calibration based on Galactic globular clusters \citep{vale+2004b}.
  Taking into account the SFH of \leoi\ by mean of synthetic CMDs, we
  corrected our measurements for the age difference between \leoi\ and
  the template GCs. This yielded a corrected distance modulus
  \corrmod, in agreement with previous optical results.
  The \vks\ colours of RGB stars were used to infer a metallicity
  distribution. We obtained a mean age-corrected metallicity \corrmh,
  consistent with recent results from spectroscopic analysis of
  \leoi\ RGB stars \citep{bosl+2007,koch+2007leo1,gull+2009spec}. A
  star-by-star comparison of metallicities derived from spectroscopy
  and \vks\ colours gave residuals generally consistent with the ages
  of the stars estimated from optical colours \citep{gull+2009spec}.
  Therefore, we argued that NIR photometry allows an accurate
  determination of the main parameters (distance, metallicity) of
  resolved galaxies provided that a guess of their SFH can be
  made. This conclusion is important for deriving distances and
  metallicities for resolved galaxies out to the Virgo cluster of
  galaxies with the future generation of NIR instruments (JWST and
  adaptive optics at Extremely Large Telescopes).

  Using our NIR photometry, we obtained a nearly complete (with the
  possible exception of heavily obscured stars), homogeneous sample of
  AGB star in \leoi, confirming the presence of some very red objects
  not detected in optical images, probably luminous thermally-pulsing
  AGB stars reddened by dust envelopes \citep[4 out of 5 in common
  with ][]{menz+2002}.
  We showed that, if the photometric data are of high precision and
  accuracy, the NIR two-colour diagram is able to discriminate between
  O-rich and C-rich AGB stars brighter than the RGB tip, and against
  the foreground Galactic dwarf stars.  Our selection criteria for C
  stars brighter than the TRGB are entirely based on NIR photometry,
  hence not affected by the various biases typical of optical methods.
  The same method can be employed to select a clean sample of red
  giant stars by significantly reducing the contamination by Galactic
  dwarf stars.
  As a result, we found \numc\ C-stars in \leoi, 6 of which are new
  candidates.  We were able to define, for the first time, a sample of
  \numo\ bona-fide O-rich AGB stars against the contaminating Milky
  Way foreground, on the basis of their NIR colours.  Our selection
  was tested (and confirmed) on a control field. From these samples,
  we derived the luminosity functions of O-type and C-type AGB stars
  in \leoi.

  The AGB stars in our \leoi\ sample were used as tracers of their
  parent intermediate-age stellar populations.  We revealed a radial
  gradient in the number of C-type AGB stars relative to RGB stars,
  which implies an increasing fraction of intermediate-age (1--3 Gyr)
  stars in the inner regions of the galaxy.  The fraction of C stars
  within an ellipse with semi-major axis $r=2\farcm5$ (186 pc), having
  ellipticity and position angle appropriate for \leoi, is larger by a
  factor of 3 than in the outer field.  This confirms the presence of
  an age gradient in the stellar populations of \leoi, as suggested by
  our spectroscopic study \citep{gull+2009spec}.  
  \referee{In contrast, the O-rich AGB stars follow the profile of RGB
    stars.}

  \referee{The NIR photometry of AGB stars in \leoi\ was compared with
    the prediction of theoretical models. As in the case of \leoii, we
    found significant discrepancies between the present TP-AGB models
    with low metallicity and the observed data, with the simulated LFs
    for C- and O-rich AGB stars not reproducing the observations. In
    particular, the predicted number of C stars fainter than the RGB
    tip is larger in our models than found by spectroscopic surveys.
    These discrepancies can be explained by uncertainties in the star
    formation histories used to calculate the model LFs or, more
    likely, by an overestimate of the lifetime of low- and
    intermediate-mass C-type stars, and/or an underestimate of their
    luminosities, in the current models.  }

\mytabbig{
r c@{:}c@{:}c c@{:}c@{:}c 
c c c c c
}{
\multicolumn{1}{c}{ID}&
\multicolumn{3}{c}{$\alpha$ (J2000)}&
\multicolumn{3}{c}{$\delta$ (J2000)}&
$B$&
$V$&
$J$&
$H$&
$\kk$\\}{
 10098	&10&08&44.58	&+12&17&22.8	&    20.96	&     19.42	&     16.80 	&     16.04	&     15.89	 	\\ 
 10505	&10&08&23.93	&+12&17&08.3	&    20.79	&     19.12	&     16.45 	&     15.64	&     15.48	 	\\ 
 11634	&10&08&45.32	&+12&16&32.6	&    20.18	&     18.65	&     16.07 	&     15.29	&     15.15	 	\\ 
 11760	&10&08&28.72	&+12&16&25.5	&    20.98	&     19.27	&     16.50 	&     15.71	&     15.55	 	\\ 
 12378	&10&08&34.49	&+12&15&55.6	&    20.99	&     19.30	&     16.55 	&     15.74	&     15.59	 	\\ 
 12596	&10&08&38.36	&+12&15&44.7	&    21.12	&     19.48	&     16.66 	&     15.83	&     15.67	 	\\ 
 12651	&10&08&33.66	&+12&15&40.9	&    21.10	&     19.33	&     16.43 	&     15.63	&     15.46	 	\\ 
 12779	&10&08&33.68	&+12&15&33.5	&    21.41	&     19.64	&     16.70 	&     15.93	&     15.79	 	\\ 
 12955	&10&08&08.01	&+12&15&18.0	&    21.12	&     19.33	&     16.43 	&     15.56	&     15.38	 	\\ 
 13430	&10&08&21.36	&+12&14&42.8	&    21.26	&     19.22	&     16.08 	&     15.21	&     15.02	 	\\ 
 13758	&10&08&35.08	&+12&14&08.4	&    20.27	&     18.82	&     16.39 	&     15.66	&     15.54	 	\\ 
 13911	&10&08&21.39	&+12&13&48.6	&    21.27	&     19.38	&     16.42 	&     15.64	&     15.47	 	\\ 
 14293	&10&08&10.91	&+12&12&40.6	&    20.50	&     19.09	&     16.70 	&     16.02	&     15.88	 	\\ 
  2555	&10&08&41.83	&+12&20&46.9	&    20.28	&     18.72	&     16.12 	&     15.32	&     15.17	 	\\ 
  2699	&10&08&24.53	&+12&20&40.1	&    20.80	&     19.08	&     16.30 	&     15.50	&     15.34	 	\\ 
  2864	&10&08&15.85	&+12&20&32.4	&    20.87	&     19.21	&     16.50 	&     15.68	&     15.51	 	\\ 
  3177	&10&08&22.06	&+12&20&20.8	&    20.53	&     19.15	&     16.79 	&     16.06	&     15.93	 	\\ 
  3478	&10&08&36.91	&+12&20&11.3	&    20.68	&     19.17	&     16.47 	&     15.64	&     15.48	 	\\ 
  3480	&10&08&33.63	&+12&20&11.0	&    20.91	&     19.44	&     16.64 	&     15.81	&     15.64	 	\\ 
  4547	&10&08&34.94	&+12&19&38.9	&    20.85	&     19.31	&     16.61 	&     15.84	&     15.71	 	\\ 
  4568	&10&08&22.04	&+12&19&37.4	&    20.93	&     19.29	&     16.67 	&     15.87	&     15.72	 	\\ 
   495	&10&08&24.88	&+12&23&59.3	&    20.47	&     19.04	&     16.48 	&     15.80	&     15.64	 	\\ 
  5078	&10&08&37.15	&+12&19&24.2	&    20.56	&     19.09	&     16.66 	&     15.86	&     15.72	 	\\ 
  5554	&10&08&26.36	&+12&19&11.1	&    20.18	&     18.87	&     16.51 	&     15.81	&     15.68	 	\\ 
  6236	&10&08&31.57	&+12&18&54.3	&    20.62	&     19.14	&     16.70 	&     15.92	&     15.79	 	\\ 
  6298	&10&08&17.15	&+12&18&51.9	&    20.71	&     19.14	&     16.61 	&     15.80	&     15.66	 	\\ 
  6458	&10&08&22.00	&+12&18&48.9	&    20.67	&     18.96	&     16.32 	&     15.51	&     15.35	 	\\ 
  6473	&10&08&28.44	&+12&18&49.0	&    20.56	&     19.20	&     16.80 	&     16.05	&     15.93	 	\\ 
   676	&10&08&12.28	&+12&23&19.4	&    20.94	&     19.28	&     16.75 	&     16.03	&     15.89	 	\\ 
  6878	&10&08&26.58	&+12&18&40.1	&    21.01	&     19.37	&     16.79 	&     16.07	&     15.90	 	\\ 
  6961	&10&08&23.26	&+12&18&38.1	&    20.77	&     19.15	&     16.54 	&     15.72	&     15.56	 	\\ 
  7216	&10&08&29.25	&+12&18&31.9	&    20.69	&     18.95	&     16.21 	&     15.41	&     15.24	 	\\ 
  7240	&10&08&39.02	&+12&18&32.2	&    21.10	&     19.44	&     16.46 	&     15.63	&     15.46	 	\\ 
  7924	&10&08&17.07	&+12&18&14.7	&    21.16	&     19.51	&     16.73 	&     15.94	&     15.79	 	\\ 
  8366	&10&08&32.61	&+12&18&05.9	&    20.73	&     19.18	&     16.66 	&     15.88	&     15.75	 	\\ 
  8476	&10&08&14.57	&+12&18&01.7	&    20.93	&     18.98	&     16.09 	&     15.24	&     15.06	 	\\ 
  8667	&10&08&20.25	&+12&17&57.8	&    19.89	&     18.13	&     15.35 	&     14.52	&     14.38	 	\\ 
  8836	&10&08&27.89	&+12&17&54.8	&    21.44	&     19.59	&     16.66 	&     15.82	&     15.66	 	\\ 
  9955	&10&08&10.75	&+12&17&24.0	&    20.95	&     19.28 & 16.67 	&     15.86	&     15.71	 	\\
}
{
  Candidate O-rich AGB stars in \leoi, as selected from the NIR
  two-colour diagram (Fig.~\ref{f:2colbox}, region 1).
}{
t:box1}{
normalsize}

\mytabbig{
r c@{:}c@{:}c c@{:}c@{:}c 
c c c c c
l}{
\multicolumn{1}{c}{ID}&
\multicolumn{3}{c}{$\alpha$ (J2000)}&
\multicolumn{3}{c}{$\delta$ (J2000)}&
$B$&
$V$&
$J$&
$H$&
$\kk$&
note\\}{
 10027	&10&08&35.28	&+12&17&24.5	&    21.50	&     19.46	&     16.21 	&     15.46	&     15.23		& ALW-18 \\ 
 10531	&10&08&19.49	&+12&17&07.1	&    21.48	&     19.53	&     16.44 	&     15.69	&     15.39		& ALW-1 \\ 
 10543	&10&08&31.03	&+12&17&07.7	&    20.86	&     18.80	&     15.74 	&     14.94	&     14.69		& ALW-14 \\ 
 10975	&10&08&27.51	&+12&16&53.9	&    21.67	&     19.47	&     16.97 	&     15.97	&     15.30		& ALW-9,v \\ 
  1208	&10&08&39.88	&+12&22&14.6	&    22.34	&     19.83	&     16.30 	&     15.29	&     14.72		& DB-C08 \\ 
 13680	&10&08&23.87	&+12&14&16.6	&    22.60	&     19.85	&     16.68 	&     15.65	&     15.04		& ALW-4 \\ 
 13795	&10&08&26.34	&+12&14&02.8	&    21.82	&     19.55	&     16.08 	&     15.25	&     14.95		& ALW-6 \\ 
 14111	&10&08&01.09	&+12&13&14.7	&  \nodata	&   \nodata	&     18.42 	&     16.59	&     15.14		& \\ 
  1478	&10&08&04.06	&+12&21&46.9	&    20.88	&     19.43	&     16.16 	&     15.42	&     15.21		& \\ 
 20915	&10&08&22.69	&+12&23&15.8	&  \nodata	&   \nodata	&     17.05 	&     15.47	&     14.34		& DB-C13 \\ 
 21484	&10&08&20.09	&+12&20&02.4	&  \nodata	&   \nodata	&     17.12 	&     15.74	&     14.83		& DB-C02 \\ 
  2158	&10&08&35.00	&+12&21&03.2	&    21.14	&     19.16	&     16.06 	&     15.26	&     14.97		& ALW-16,v \\ 
  2641	&10&08&29.53	&+12&20&43.0	&    21.22	&     19.39	&     16.43 	&     15.65	&     15.38		& ALW-13 \\ 
  3493	&10&08&18.08	&+12&20&09.2	&    21.52	&     19.32	&     16.27 	&     15.39	&     15.05		& \\ 
  3554	&10&08&16.60	&+12&20&07.0	&    21.13	&     19.27	&     16.50 	&     15.72	&     15.51		& ALW-19 \\ 
  4137	&10&08&28.50	&+12&19&48.6	&    21.02	&     19.58	&     16.10 	&     15.20	&     14.80		& ALW-11,DB-C12 \\ 
  4458	&10&08&48.36	&+12&19&42.0	&    21.00	&     19.37	&     16.83 	&     16.12	&     15.93		& \\ 
  4523	&10&08&12.89	&+12&19&37.9	&    22.72	&     19.99	&     15.88 	&     14.96	&     14.51		& DB-C11 \\ 
  4951	&10&08&24.82	&+12&19&26.9	&    21.09	&     19.09	&     15.79 	&     14.94	&     14.62		& ALW-7 \\ 
  6101	&10&08&27.30	&+12&18&57.3	&  \nodata	&   \nodata	&     16.99 	&     15.31	&     14.09		& M02-B,v \\ 
  6103	&10&08&25.62	&+12&18&57.1	&  \nodata	&   \nodata	&     16.04 	&     15.08	&     14.58		& DB-C10,v \\ 
   626	&10&08&28.23	&+12&23&30.1	&    21.61	&     19.52	&     16.34 	&     15.56	&     15.32		& \\ 
  6343	&10&08&29.28	&+12&18&51.6	&  \nodata	&   \nodata	&     17.89 	&     16.03	&     14.55		& M02-A,v \\ 
  6598	&10&08&32.34	&+12&18&46.2	&  \nodata	&   \nodata	&     15.56 	&     14.59	&     14.10		& DB-C07,v \\ 
  7007	&10&08&34.71	&+12&18&37.6	&    21.08	&     19.18	&     15.79 	&     14.95	&     14.63		& ALW-15 \\ 
  7013	&10&08&20.65	&+12&18&36.5	&  \nodata	&   \nodata	&     15.60 	&     14.54	&     13.97		& DB-C03 \\ 
  7095	&10&08&11.71	&+12&18&33.5	&    23.27	&     20.04 &     15.75 	&     14.70	&     14.12		& DB-C04 \\ 
  7276	&10&08&25.29	&+12&18&30.2	&    20.63	&     19.39	&     15.72 	&     14.78	&     14.30		& ALW-8,DB-C06,v \\ 
  7450	&10&08&22.55	&+12&18&25.9	&    21.40	&     19.22	&     15.85 	&     14.92	&     14.52		& ALW-5,DB-C05 \\ 
  7965	&10&08&15.76	&+12&18&13.7	&    21.15	&     19.32	&     16.73 	&     16.01	&     15.82		& \\ 
  8336	&10&08&41.20	&+12&18&07.1	&    18.77	&     18.73 &     17.30 	&     15.87	&     14.61		& M02-D \\ 
  8717	&10&08&22.25	&+12&17&57.0	&  \nodata	&   \nodata &     17.18 	&     15.22	&     13.57		& M02-C,v \\ 
  9344	&10&08&19.96	&+12&17&41.4	&    21.69	&     19.54	&     15.65 	&     14.79	&     14.44		& ALW-2,DB-C01 \\ 
  9745	&10&08&27.68	&+12&17&31.8	&    21.43	&     19.46	&     16.47 	&     15.73	&     15.43		& ALW-10 \\ 
  9951	&10&08&21.76	&+12&17&25.0	&  \nodata	&   \nodata	&     16.11 	&     15.17	&     14.72		& ALW-3,DB-C09 \\ 
 7125 &  10&08&30.08 &  +12&18&34.2  &  21.11 &  19.59  & 17.11 & 16.45 &  16.26 &  ALW-12   \\
  11242 & 10&08&40.39 & +12&16&46.0   &  21.52 &  19.88  & 17.42 &  16.77 &  16.59 &  ALW-17,C? \\
}
{
Candidate C-rich AGB stars in \leoi. All stars were selected from the
NIR two-colour diagram (Fig.~\ref{f:2colbox}, region~4), with the
exception of stars 7125 and 11242 which are spectroscopically
identified C stars fainter than the TRGB.  Cross-identifications with
the surveys of \citet[][ALW]{azzo+1986}, \citet[][DB]{demebatt2002}
and \citet[][M02]{menz+2002} are given. Probable variable stars in M02
are labelled ``v".
}{
t:box4}{
normalsize}

\section*{Acknowledgements}

We thank M. Riello from the CASU Astronomical Data Centre
for helpful comments and support with the WFCAM pipeline. We also
thank Y. Momany for help and fruitful discussions in the course of
this project. We acknowledge funding by the INAF PRIN07 project CRA
1.06.10.03.  EVH wishes to acknowledge the hospitality of the Joint
Astronomical Centre at Hilo where this paper was partly written.
This publication made use of data products from the Two Micron All Sky
Survey, which is a joint project of the University of Massachusetts
and the Infrared Processing and Analysis Center/California Institute
of Technology, funded by the National Aeronautics and Space
Administration and the National Science Foundation.

\label{lastpage}

\end{document}